\documentclass[%
 reprint,
 amsmath,amssymb,
 aps,
 floatfix,
 superscriptaddress
]{revtex4-2}

\usepackage{graphicx}
\usepackage{dcolumn}
\usepackage{bm}
\usepackage{hyperref}
\usepackage{comment}
\usepackage{upgreek}
\usepackage{notes2bib}
\usepackage{xcolor}

\newcommand{\porder}{\langle\left|\langle e^{i\theta}\rangle_S\right|\rangle_t}
\newcommand{\pauto}{\langle\bm{n}_p(t)\cdot\bm{n}_p(t+\tau)\rangle_t}
\newcommand{\mum}{\upmu\textrm{m}}
\begin{document}
\preprint{APS/123-QED}

\title{Algebraic correlations and anomalous fluctuations in ordered flocks of \\Janus particles fueled by an AC electric field}

\author{Junichiro Iwasawa}
\email{jiwasawa@ubi.s.u-tokyo.ac.jp}
 \affiliation{Department of Physics, The University of Tokyo, Hongo 7-3-1, Tokyo, 113-0033, Japan}
\author{Daiki Nishiguchi}
 \email{nishiguchi@noneq.phys.s.u-tokyo.ac.jp}
\affiliation{Department of Physics, The University of Tokyo, Hongo 7-3-1, Tokyo, 113-0033, Japan}
\author{Masaki Sano}
\email{sano.masaki@sjtu.edu.cn}
\affiliation{Institute for Natural Sciences, School of Physics and Astronomy, Shanghai Jiao Tong University, Shanghai 200240, China}
\affiliation{Universal Biology Institute, Graduate School of Science, The University of Tokyo, Hongo 7-3-1, Tokyo, 113-0033, Japan}
\date{\today}

\begin{abstract}
We study the polar collective dynamics of Janus colloidal particles fueled by an AC electric field.
When the density is high enough, the polar interactions between the particles induce a polar orientationally ordered state which exhibits features reminiscent of the Vicsek model such as true long-range order and giant number fluctuations.
Independent measurements of the polarity and velocity at the single particle level allowed us to investigate the single particle dynamics within the ordered state.
We discovered theoretically-unaddressed statistical properties of the ordered state such as the asymmetric relation of polarity and velocity, enhanced rotational diffusion stronger than in the disordered state, and an algebraic auto-correlation of the polarity.
Our experimental findings, at the crossroad of the Vicsek physics and the Active Brownian Particles physics, shed light on the so-far-unexplored physics arising from the interplay between the polarity and the velocity.
\end{abstract}
\maketitle

\section{Introduction}
Active matter systems consume and dissipate energy at the level of its local units to generate systematic motion based on their internal degrees of freedom (e.g. polarities) \cite{ramaswamy2010,marchetti2013,ramaswamy2017,chate2020}.
After decades of extensive research, it is now known that active matter systems can exhibit a wide range of phenomena such as phase separation \cite{tailleur2008, theurkauff2012,palacci2013,linden2019}, active chain formation \cite{nishiguchi2018a,harder2018}, 
active turbulence \cite{zhang2010,wu2017,nishiguchi2018b,nishiguchi2020}, etc.
Among these topics, orientational order in active matter has gained considerable attention since the introduction of the Vicsek model in 1995, where pointwise polar particles move at constant speed toward their polarity and try to align their polarities with local neighbors under the presence of noise \cite{vicsek1995}.
Theoretical work concerning the Vicsek-style models has been done including particle, kinetic and hydrodynamic levels \cite{tonertu95, tonertu98, ramaswamy2003,chate2004,bertin2006,chate2008,bertin2009,toner2012,benoit2019}.
In particular, the phenomenological hydrodynamic equations by Toner, Tu, Ramawsamy and their coworkers have predicted the existence of true long-range order and giant number fluctuations (GNF) for the orientationally ordered state \cite{tonertu95,tonertu98,ramaswamy2003,toner2005,toner2012}.
In these theoretical and numerical studies, each particle's polarity is assumed to be parallel to its velocity.
In experimental systems, on the other hand, particles do not necessarily move toward their polarities due to thermal and/or athermal noise and inter-particle (hydrodynamic, excluded volume, or electrostatic) interactions (e.g. systems of elongated bacteria \cite{nishiguchi2017}, vibrated polar disks \cite{dauchot2013}, motility assay \cite{tanida2020}, and active rolling colloids \cite{bricard2013,bartolo2018}). 

\begin{figure}[tb]
\includegraphics[width=\columnwidth]{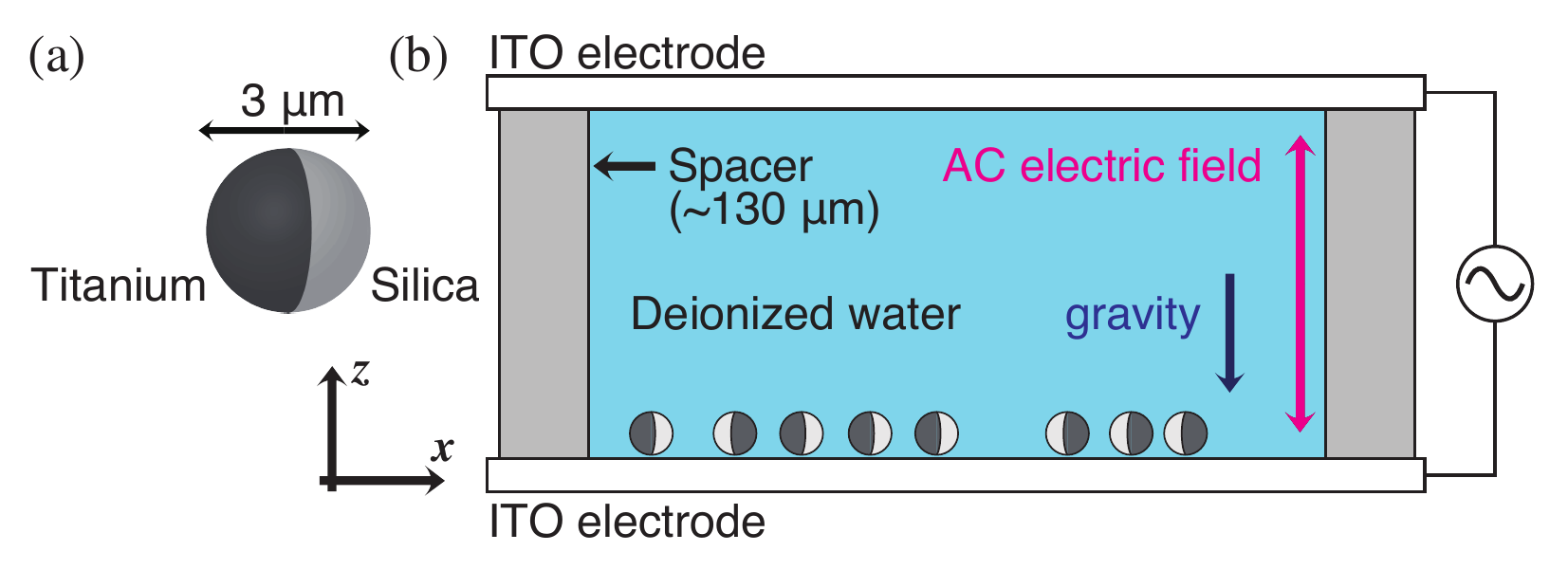}
\caption{\label{schematic}Schematic image of the experimental system. (a) A silica based Janus particle with a metallic titanium hemisphere. (b) The suspension of Janus particles were sandwiched by two ITO coated electrodes. The particles sediment to the surface of the ITO electrode which leads to quasi-two-dimensional motion on the $xy$-plane.}
\end{figure} 

One of the minimal models for investigating such active systems whose particle velocities do not necessarily coincide with their polarities is the Active Brownian Particles (ABP) model.
In the ABP model, particles self-propel with a constant self-propulsive force under rotational/translational noise. The particles interact only by volume exclusion and do not exhibit any aligning interactions~\cite{callegari2019,poncet2020}.
In spite of the absence of alignment, intriguing collective phenomena such as motility-induced phase separation have been reported for ABP~\cite{cates2015}.
However, it still remains unclear what lies between non-aligning ABP-like systems and the aligning Vicsek-style models.
Active colloidal systems similar to the ABP model have been experimentally realized and studied by using e.g. auto-catalytic colloidal particles in hydrogen peroxide and rolling colloids under a DC electric field. 
However, active particles fueled by chemicals do not exhibit ordered states reminiscent of those in Vicsek-style models so far~\cite{theurkauff2012,palacci2013,ginot2018}.
In the case of rolling colloids, because the particles do not possess intrinsic polarity and that their polarities are only electronic and invisible, in-depth study of the interplay between the polarity and velocity in the ordered state is not possible~\cite{bartolo2018}.
Thus, our understanding of how polarity and velocity interact with each other and how they behave in orientationally long-range ordered colloidal systems with hydrodynamic interactions remains elusive.
For example, in the systems of elongated bacteria and motility assay, although true long-range order has been observed, particles could only be detected in a coarse grained manner \cite{nishiguchi2017,tanida2020}.

Janus particles \cite{deGennes1992}, which are colloidal particles with two distinct hemispheres, offer a great platform for investigating active matter experimentally \cite{jiang2010,theurkauff2012, nishiguchi2015, granick2016,mano2017,nishiguchi2018a,linden2019,poncet2020}.
Among these Janus particles, the ones fueled by an AC electric field are especially suitable for analysis from the view point of statistical physics because they constantly convert energy into motion without running out of energy sources and thus, allow long enough measurements for statistical analysis \cite{suzuki2011,nishiguchi2015,mano2017,nishiguchi2018a,linden2019,poncet2020}, which is not the case for auto-catalytic particles due to the consumption of chemical fuels.
Under an AC electric field, Janus particles display two different mechanisms of swimming depending on the field frequency: induced charge electrophoresis (ICEP) at low frequency \cite{squires2004,squires2006,gangwal2008}; self-dielectrophoresis (sDEP) at high frequency \cite{boymelgreen2012,boymelgreen2016}.
Importantly, the two swimming mechanisms induce motion from the dielectric hemisphere to the metallic hemisphere \cite{gangwal2008,suzuki2011,nishiguchi2015,granick2016,nishiguchi2018a,linden2019,poncet2020} or vice versa \cite{suzuki2011,boymelgreen2016,granick2016,nishiguchi2018a}, which allows us to define and observe each particle's polarity based on the location of its metallic hemisphere \cite{nishiguchi2018a,linden2019}.
Not only the swimming mechanisms but also the interactions between particles depend on the field frequency and ion concentration, which lead to various types of collective behavior such as a turbulent phase \cite{nishiguchi2015}, a swarming state with vortices \cite{granick2016}, flagella-like movement \cite{nishiguchi2018a} and active clustering \cite{linden2019}.
Thus, Janus particles provide us an ideal venue for experimentally studying the collective dynamics of ordered systems while keeping track of polarity at the single particle level.

\begin{figure*}[t!]
\includegraphics[width=1.7\columnwidth]{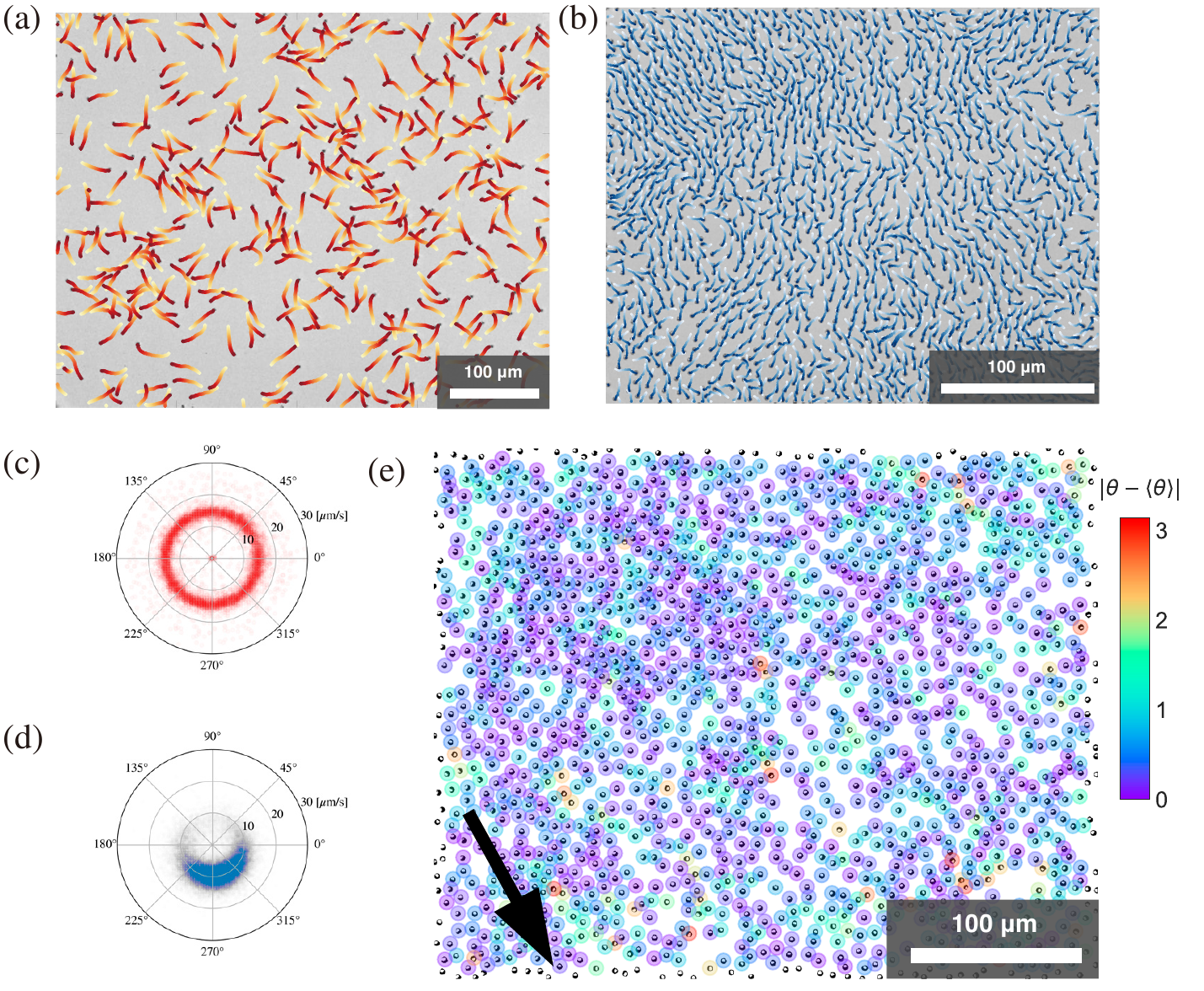}
\caption{\label{ordered_janus} Dynamics of Janus particles in the disordered and ordered states. 
(a), (b) Trajectories of the particles in the disordered state (three seconds) and the ordered state (two seconds), respectively. The bright and dark colors indicate the early and late times of each trajectory, respectively.
(c), (d) Velocity distribution of the particles in the disordered (9909 particles) and ordered state (40205 particles), respectively \cite{supplement}.
The mean speeds for the disordered and ordered state were $\bar{v}_d=14.6\ \upmu$m/s and $\bar{v}_o = 8.5\ \upmu$m/s, respectively.
(e) Binarized experimental snapshot of the Janus particles in the ordered state.
Each particle is colored based on the deviation of its polarity from the global order $\left|{\theta}-\langle\theta\rangle\right|$ (radian).
The black arrow shows the direction of global order $\langle\theta\rangle$.}
\end{figure*}

Here in this paper, we perform experiments on the quasi-two dimensional system of active Janus particles fueled by an AC electric field. 
We specifically use the AC field frequency regime where Janus particles swim by sDEP and exhibit a flocking phase.
Through the extraction of both the polarity and velocity for each particle, we show that this flocking phase of Janus particles can exhibit true long-range orientational order.
This ordered state of Janus particles also exhibited statistical features such as GNF and algebraic correlations for orientation fluctuations which are reminiscent of the predictions of Toner and Tu~\cite{tonertu95,tonertu98,toner2012}.
We further investigate the space-time correlations of density fluctuations leading to an estimate of dynamical and anisotropy exponents for the ordered state of Janus particles.
In addition, through single particle measurements, we show that the particles counterintuitively exhibit larger diffusion in the ordered state than that in the disordered state for short time scales.
We suggest that this anomalous enhanced diffusion stems from the coupling between particle polarity and velocity through computations of the cross-correlation function, which is a dismissed feature in Vicsek-style models.
Our work, through the observation of both velocities and polarities, unravels hitherto unexplored statistical properties within orientationally ordered active colloidal systems.

\begin{figure}[bt]
\includegraphics[width=\columnwidth]{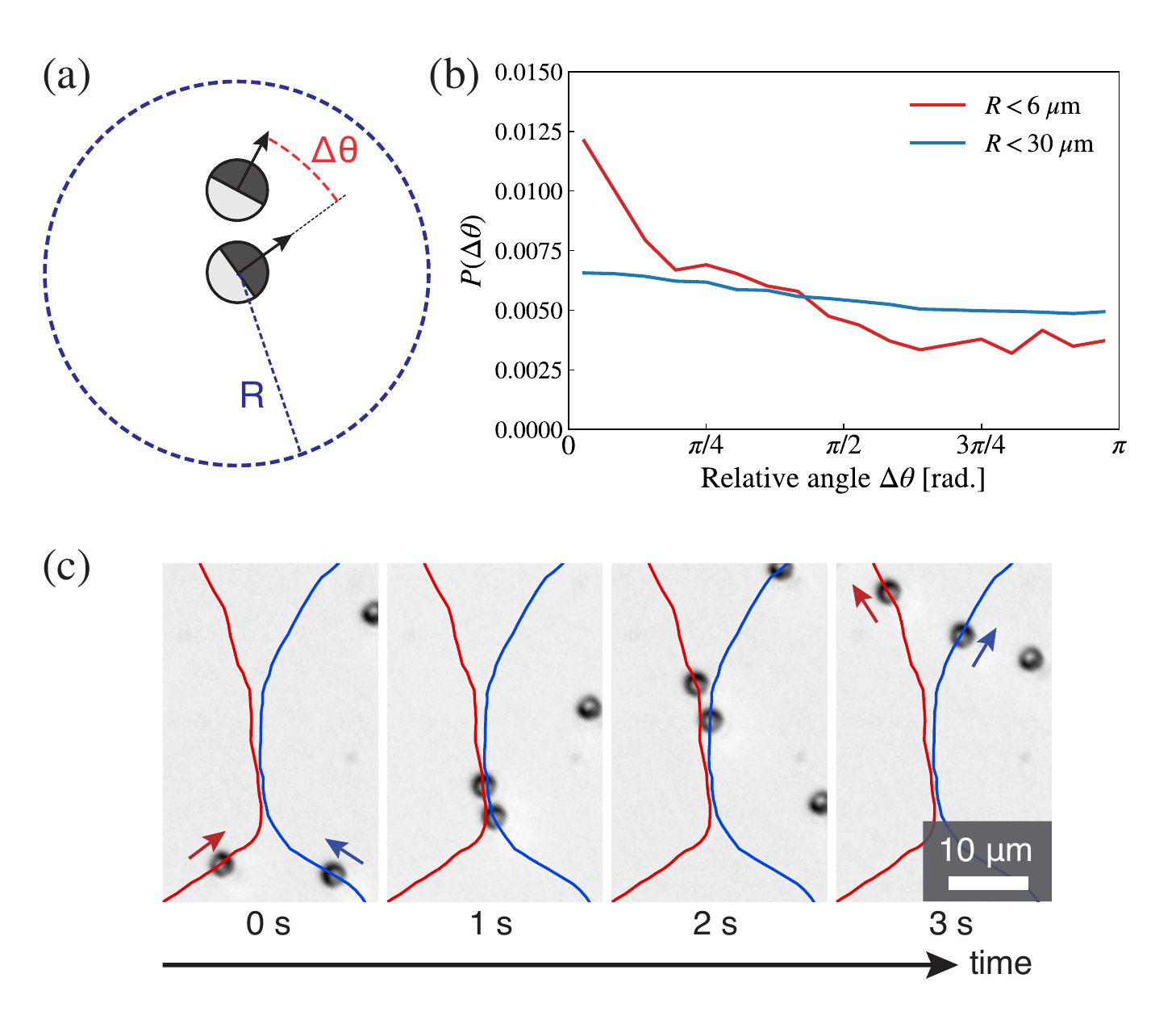}
\caption{\label{binary_int} Statistics of binary interactions of Janus particles. (a) Schematic picture describing the relative angle of particle pairs. 
(b) The relative polarity angle $\Delta \theta$ distribution of Janus particles in a low density situation. $R$ denotes the distance of two isolated Janus particles. 
(c) Typical trajectories of two Janus particles when they are in close distance. }
\end{figure} 

\section{Experimental System}
Our experimental system is composed of silica based Janus particles with 3.17 $\upmu$m diameter \cite{nishiguchi2018a}. 
To create distinct hemispheres, 35 nm of titanium followed by 15 nm of silica were deposited to a monolayer of silica colloidal particles (Bangs Laboratories, Inc., SS05N) using electron beam deposition and thermal evaporation, respectively. 
The additional 15 nm layer of silica was deposited to suppress oxidization and the subsequent changes of electrical properties and color (from black to transparent) of the Ti surface.
After deposition, the particles were suspended into deionized water and washed by sonication to prevent particles from forming clusters. 
This washing procedure was repeated three times after waiting for 30 -- 60 minutes for particle sedimentation. 
The washed Janus particles were suspended in deionized water and then sandwiched by two 25-nm-silica-deposited indium tin oxide (ITO) coated electrodes (Mitsuru Optical Co. Ltd.), separated by 130 $\upmu$m thick double sided tape. 
Before applying the AC electric field, we waited for a few minutes to ensure that the particles have sedimented to the surface of the bottom electrode, which leads to a quasi-two dimensional system of Janus particles (Fig.~\ref{schematic}). 
Accordingly, the induced-charge electro-osmotic flow caused by the electric field lets the particle's polarity turn to the horizontal $xy$-plane, leading to confined motion in the horizontal 2D-plane \cite{kilic2011}. 
A sinusoidal AC electric field with $f = 1$ MHz frequency and 16 ${\rm V_{pp}}$ voltage was applied for all experiments, where ${\rm  V_{pp}}$ denotes the peak-to-peak voltage.
In this frequency regime, it is known that Janus particles tend to move toward its metallic hemisphere, which contradicts with the ICEP theory \cite{squires2004, squires2006, gangwal2008}. This was first observed in \cite{suzuki2011}, and is now understood as sDEP \cite{boymelgreen2012, boymelgreen2016}.
It should be noted that the ordinary electrophoretic effects caused by the negative charge of the silica particles could be negligible in our system since we use an AC electric field. 
Therefore, ordinary electrophoresis alone cannot compete with the ICEP/sDEP mechanism.
Experiments were performed under two different density conditions: the low density regime where $\rho\simeq0.78$ particles/$1000\ \upmu$m$^2$, and the high density regime where $\rho\simeq13$ particles/$1000\ {\rm \upmu m}^2$.

The observations of Janus particles were done using an inverted microscope (Nikon ECLIPSE TE2000-U) with objective lens ((Nikon Plan Fluor ELWD, 20$\times$, NA=0.45 and 40$\times$, NA=0.60 for the low and high density regime, respectively).
A halogen lamp was used to illuminate the particles and a green filter was inserted in the light path before the sample to increase the contrast of the obtained images.
Under this condition, the polarity of each Janus particle can be distinguished using its dark-gray hemisphere indicating the titanium side, and the light-gray hemisphere indicating the silica-only side (for example see Fig.~\ref{binary_int}(a), Supplemental Movie 2, 3). 
The dynamics were recorded at 10 fps and 15 fps for the low and high density regime, respectively, using a CMOS camera (Baumer LXG80, $3000\times 2400$ pixels $\simeq 430 \mum \times 340 \mum$ ($40\times$)). This allowed us to capture the large scale statistics of the particles while detecting the polarities and velocities of each particle at the same time.

Image analysis was performed using Trackpy \cite{trackpy} and ImageJ \cite{imagej}.
Specifically, the polarities of the particles were detected by taking the difference of the centroid and center of mass of image intensity for each particle using the built-in algorithms in ImageJ \cite{supplement}.
The duration of the analyzed movies were 360 s and 212 s for the low and high density regime, respectively.

\begin{figure*}[!tb]
\includegraphics[width=2\columnwidth]{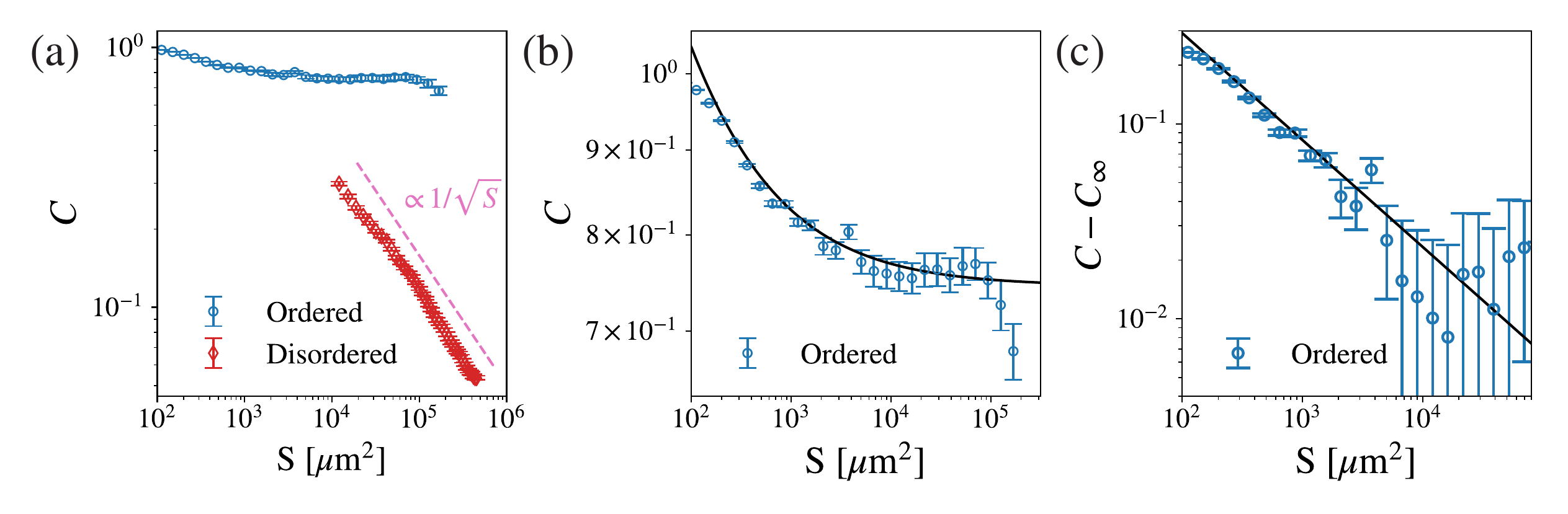}
\caption{\label{order_param} The polar order parameter for the ordered and disordered states.
(a) Log-log plot of the polar order parameter $C$ vs $S$, the area of the ROI. The blue and red points show the order parameter for the ordered and disordered state, respectively. 
(b) The same data as in (a) for the ordered state in a magnified range (log-log). 
(c) The same data as in (b) with the asymptotic value $C_\infty = 0.74$ subtracted (log-log).
The solid black line shows the nonlinear fit: $C = C_\infty + k S^{-\gamma/2}$ with $C_\infty = 0.74$, $k = 3.1$, and $\gamma/2 = 0.53$. 
The data for ROIs with $ S>10^{5}\ \upmu$m$^2$ have been excluded from the nonlinear fitting due to the existence of counterflows in the system.
Error bars in (a), (b), (c): standard error.}
\end{figure*}

\section{Results}
\subsection{Binary polar interaction and following motion}
To probe the interaction of the Janus particles in our system, we first measured the relative polarity distribution $P\left(\Delta\theta\right)$ for all Janus particle pairs in the low density regime (Fig.~\ref{ordered_janus}(a)). 
To examine the local interactions of the particles, particle pairs within 6 $\upmu$m and 30 $\upmu$m distance were extracted, respectively, and the relative polarity angle $\Delta\theta$ was measured for each pair (Fig.~\ref{binary_int}(a)(b)) \cite{supplement}. 
The results show that particles within 6 $\upmu$m range had a narrower distribution with a peak near $\Delta\theta=0$, which suggests the existence of polar interaction within the Janus particles.
Indeed, when observing two Janus particles within a short distance, following motion where one particle tends to follow the other particle could often be observed (for example, see Fig.~\ref{binary_int}(c)). 

The observed polar interaction and following motion can be understood as a result of the balance of electrostatic and hydrodynamic interactions between the particles.
Although the interaction mechanism of Janus particles under an AC electric field is still under discussion \cite{granick2016, nishiguchi2018a}, our previous study experimentally validated the existence of polar interaction induced by the effective electrostatic dipole-dipole interaction between two particles both in the ICEP and the sDEP regimes \cite{nishiguchi2018a}.
Note that, although our study in the sDEP regime uses a similar AC field frequency as in \cite{nishiguchi2018a}, the smaller ion concentration in this experiment than in our previous study generates a longer screening length (Debye length) $\lambda_D \propto 1/\sqrt{c}$, where $c$ is the ion concentration \cite{nishiguchi2018a}. 
The slip velocity $\bm{u}_s$ generated in the electric double layer scales as $\bm{u}_s \propto \lambda_D$, following from the Helmholtz-Smoluchowski formula \cite{squires2004}. 
This results in a weak but relatively stronger pusher-type flow-field than in our previous study \cite{nishiguchi2018a} which prevents the particles from forming active chains in the current setup.
It should also be noted that flow fields calculated based on pusher-type flow fields also show following motion due to the reorientation of the following particle, which is consistent with our observations \cite{lauga2009,stocker2012}.
This synergy of electrostatic and hydrodynamic interactions leads to polar alignment, allowing us to observe the ordered dynamics of Janus particles as we see below.

\begin{figure*}[!t]
\includegraphics[width=2\columnwidth]{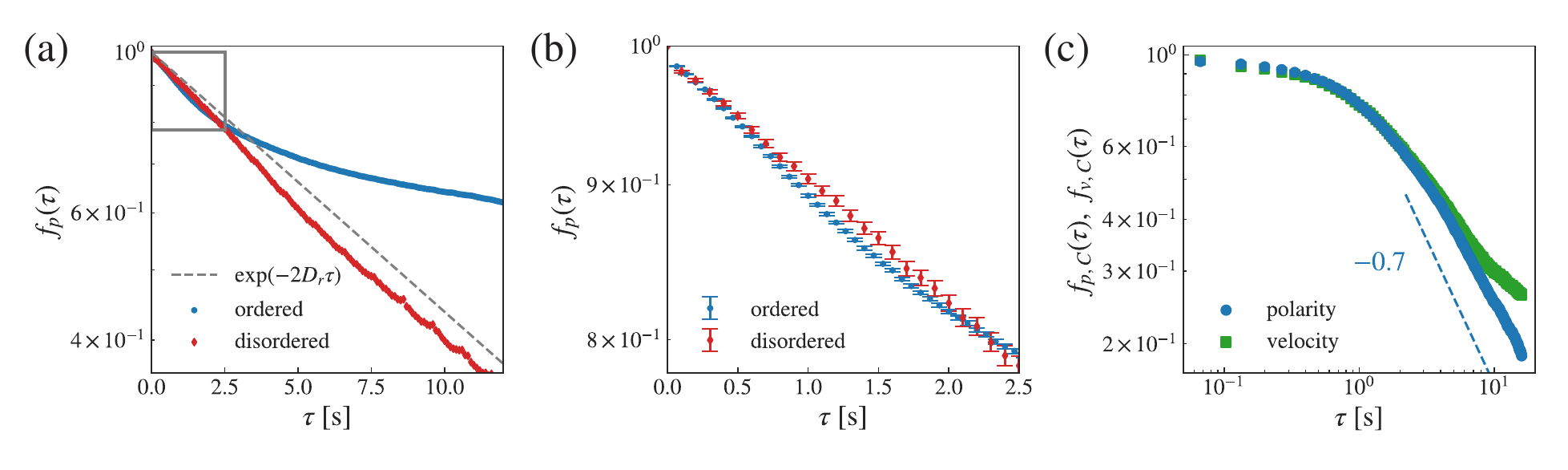}
\caption{\label{auto_corr} Auto-correlation functions. 
(a) Polarity auto-correlation function in the ordered (red) and disordered state (blue) on a semi-log scale. $\tau$ denotes the time delay. 
The dashed line shows $\exp{\left(-2D_r \tau\right)}$ where $D_r \simeq 1/24.2$ s$^{-1}$ is the rotational diffusion constant for a $3.17\ \upmu\textrm{m}$ diameter particle derived from the Einstein-Stokes relation.
(b) The same data as in (a) in a magnified view which is shown as a gray box in (a) (semi-log). 
Error bars show the standard error. (c) The connected auto-correlation functions for polarity:$f_{p,C}(\tau)$ and velocity: $f_{v,C}(\tau)$ in the ordered state.
The blue dashed line with a slope: $-0.7$ is a guide for the eye.}
\end{figure*}

\subsection{True long-range order}
Janus particles exhibited distinct global patterns when we changed particle density.
While Janus particles exhibited uncorrelated disordered motion in the low density regime (Fig.~\ref{ordered_janus}(a), (c), Supplemental Movie 1), orientationally-ordered coherent motion was observed in the high density regime (Fig.~\ref{ordered_janus}(b), (d), (e), Supplemental Movie 3, 4) due to the local polar interaction. This density dependency of Janus particles is consistent with the previous experimental results of \cite{granick2016}. Note, if we compare the Janus particle system with Vicsek-style models, the density and the noise amplitude should be the basic factors which control the transition from the disordered state to the ordered state. Although the effective noise amplitude, namely the P\'{e}clet number, could be adjusted by changing the particle's velocity via the AC electric field voltage, we have not been able to investigate this in the current study.
This is because the particles easily adhere to the surface of the ITO electrode at higher voltages forbidding long time observation.
Ideally, the effective noise amplitude could also be changed by using particles with different diameters, but it would be nearly impossible to continuously change the noise amplitude in this case.
It should also be noted that the global orientation of the system varied each time we conducted the experiment, and thus, is not deterministic. This is analogous to the setting of the Vicsek-style models where the global orientation results from spontaneous symmetry breaking of the continuous rotational symmetry of the system.

To quantitatively assess the properties of orientational order in the low and high density regimes of the system, we computed the polar order parameter,
\begin{equation}
    C = \porder,
\end{equation}
for different sizes of rectangular (5:4) regions of interest (ROIs) taken in spatially-uncorrelated locations (see section \ref{sup_calculation_order_parameter} in the supplemental material for details), where $\theta$ is the particle polarity and $\langle\cdot\rangle_S$ and $\langle\cdot\rangle_t$ denotes the average within the same ROI and statistically independent time frames, respectively (Fig.~\ref{order_param}, \cite{supplement}).
In the low density regime, the polar order parameter decays as a power law $C\propto 1/\sqrt{S}$ (Fig.~\ref{order_param}(a)). 
This is consistent with the case with $N$ randomly oriented particles. 
In contrast, in the high density regime, we can observe a decay slower than a power law within $S \sim 10^5\ \upmu$m$^2$. 
This asymptotic behavior characterizes the true long-range order, which was previously reported in Vicsek-style models \cite{chate2020} and in the bacterial experiment \cite{nishiguchi2017} where the orientational order parameter converges to a positive finite value.
Note, the decay slower than a power law in our Janus particle system is observed for about 2.8 decades, which is larger than the range reported in the long-range ordered phase of bacteria \cite{nishiguchi2017} and even comparable to the range reported in numerical simulations (see Fig. 4a of \cite{chate2020}).
To confirm the existence of true long-range order, we performed nonlinear fitting for the polar order parameter at $S<10^5\ \upmu\textrm{m}^2$, using the \texttt{curve\_fit} function from the scipy package \cite{scipy}. 
The data was in good fit with an algebraic convergence to a finite value $C-C_\infty \sim S^{-\gamma/2}$ where $C_\infty = 0.74$ and $\gamma/2 = 0.53$ (see the inset of Fig.~\ref{order_param}(b)), which indicates true long-range order in the system.
It should be noted that recent simulations showed $\gamma =0.64$ for the Vicsek model \cite{chate2020}.
At the same time, however, the decay of the order parameter deviates from this trend of algebraic convergence at $S>10^5\ \upmu\textrm{m}$. 
This is caused by the existence of counter-flows of the particles (see supplemental movie 3, 4) which is unavoidable due to the closed boundary conditions and the number conservation of the particles unless we elaborate an effective periodic boundary condition by e.g. a racetrack-like confinement \cite{bricard2013}. However, in the case of our electrically driven Janus particles, micro-structures fabricated by e.g. photoresist distort the electric field, which leads to unwanted complex 3D electroconvection and destroys the two-dimensionality of the system. Therefore, we opted for using a simple system by sandwiching as large a droplet of the suspension of Janus particles as possible between the two electrodes.
Overall, the finite size scaling analysis assures that the Janus particles exhibit polar collective motion with true long-range order.

\subsection{Interplay of polarity and velocity}
To investigate the detailed dynamics of single particle trajectories, we computed the polarity auto-correlation function $f_p(\tau)$ for the disordered and ordered states, respectively. 
Here, 
\begin{equation}
    f_p(\tau) = \pauto,
\end{equation}
where $\bm{n}_p(t)$ denotes the polarity for the particle at time $t$, and $\langle\cdot\rangle_t$ takes the average for particle trajectories longer than 24 seconds (Fig.~\ref{auto_corr}(a)(b)).
We can first observe that the polarity auto-correlation decays exponentially in the disordered state.
At small time scales, this decay was slightly faster than the theoretical estimation for an Active Brownian particle $f_p(\tau) = \exp{(-2D_r\tau)}$ \cite{jiang2010,callegari2019}, where $1/D_r\sim24.2~{\rm s}$ is the rotational diffusion constant (Fig.~\ref{auto_corr}(a)) from the Einstein-Stokes relation, which is due to the polarity detection errors \cite{poncet2020}.
We thus focus on the time scale $\tau = \text{0--12}~\textrm{s}$, where the decay of the auto-correlation exhibited a slope of $21.15\pm 0.08~\textrm{s}$, close to that of the theoretical estimation.

\begin{figure}[tb]
\includegraphics[width=\columnwidth]{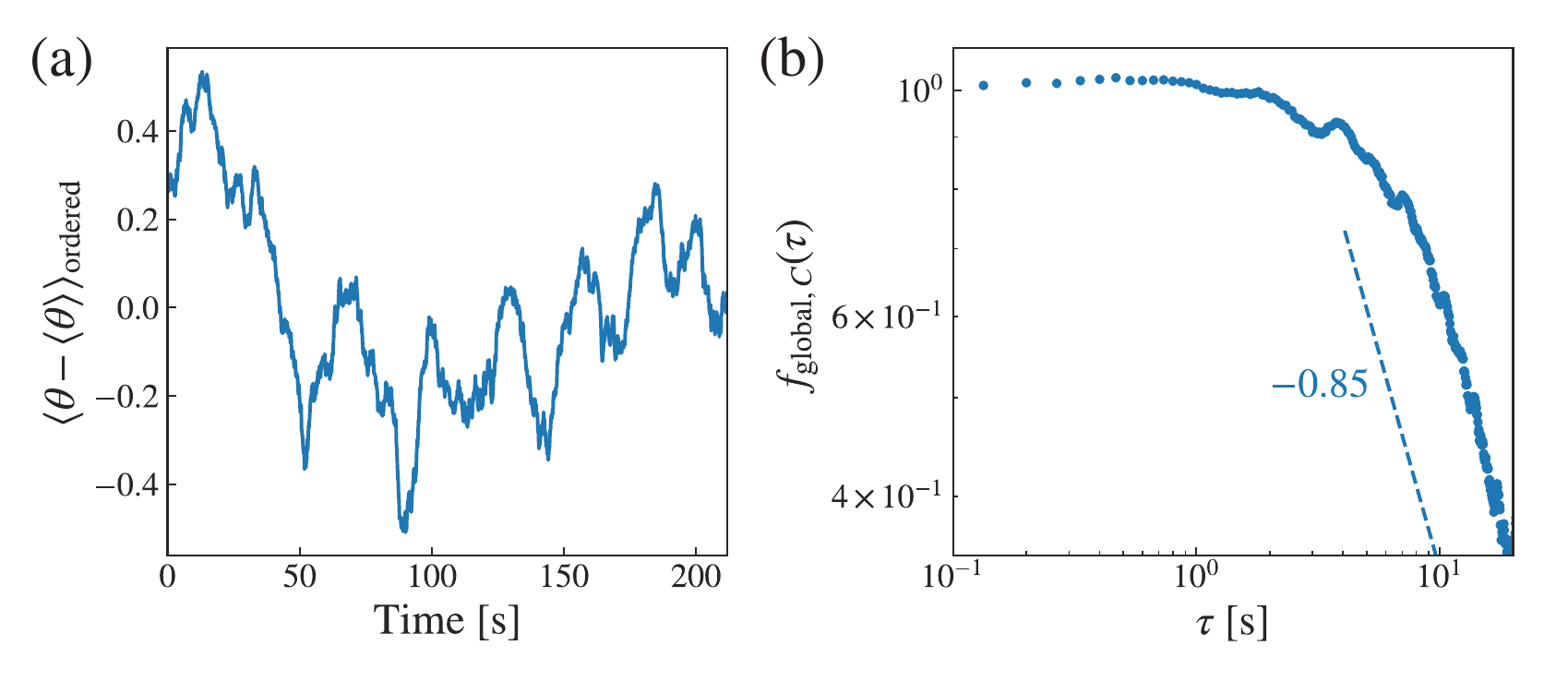}
\caption{\label{global-order} Fluctuations of the mean polarity in the ordered region. 
(a) Time series of the fluctuations of global order $\langle\theta-\langle\theta\rangle\rangle_{\rm ordered}$ where $\langle\cdot\rangle_{\textrm{ordered}}$ 
takes the average within the highly ordered region ($S<10^5~\upmu$m).
(b) Auto-correlation of the polarity fluctuations from global order 
$f_{{\rm global},C}(\tau) = \langle\Delta\tilde{\bm{n}}_p(t)\cdot\Delta\tilde{\bm{n}}_p(t+\tau)\rangle$,
where
$\Delta \tilde{\bm{n}} _{p}(t) = \langle\bm{n}_p - \langle\bm{n}_p\rangle_{t,\textrm{ordered}}\rangle_{\textrm{ordered}}$.
The blue dashed line with a slope: -0.85 is a guide for the eye.
}
\end{figure} 

In the ordered state, on the other hand, $f_p(\tau)$ exhibited a slow power-law-like decay in the long time scale ($\tau > 1$~s, Fig.~\ref{sup_autocorr}).
Since particle polarities in the ordered state tend to align to the global order, we suspected that this power-law behavior of $f_p(\tau)$ is related to the fluctuations of the global order.
To assess this hypothesis, we compared the connected polarity auto-correlation function in the ordered state with that of the global order (Fig.~\ref{global-order}).
In Fig.~\ref{auto_corr}(c), we show the connected polarity (velocity) auto-correlations,
\begin{equation}
f_{p,C}(\tau) = \langle\Delta\bm{n}_p(t)\cdot\Delta\bm{n}_p(t+\tau)\rangle_t,
\end{equation}
\begin{equation}
f_{v,C}(\tau) = \langle\Delta\bm{n}_v(t)\cdot\Delta\bm{n}_v(t+\tau)\rangle_t,
\end{equation}
where $\Delta\bm{n}_{p,v}(t) = \bm{n}_{p,v}(t)-\langle\bm{n}_{p,v}(t)\rangle$, and $\langle\bm{n}_{p,v}(t)\rangle$ is the direction of global order defined by polarity or velocity, respectively.
A power-law-like decay with a slope close to $-0.7$ for $f_{p,C}(\tau)$ could be observed for both the polarity and velocity auto-correlations.
Interestingly, the connected auto-correlation of the global order 
$f_{{\rm global},C}(\tau)$ \cite{supplement}
also showed a power-law-like decay with a slope of -0.85, close to that of $f_{p,C}(\tau)$ (Fig.~\ref{global-order}(b)).
Although there is, to the best of our knowledge, no theory predicting the slope for these power law decays in global order and polarity (velocity) auto-correlations,
these observations support our hypothesis that the power-law behavior of the polarity auto-correlations is a consequence of the fluctuations in global order.
However, it should be noted that in a finite system, the direction of global order diffuses such that the auto-correlation function decays exponentially \cite{benoit2019}. 
Therefore, we should expect that the power-law behavior of $f_{{\rm global},C}(\tau)$ that we observe is a transient to an exponential decay. 
Indeed, when we focus on longer time scales ($\tau>10$ s), we can observe a decay faster than a power-law for $f_{{\rm global},C}(\tau)$ (Fig.~\ref{sup_global_order}).
Although the limited length scale of our experimental system hinders the long time behavior of single particle trajectories, we expect that the connected polarity auto-correlation function $f_{p,C}(\tau)$ will also show a crossover to an exponential decay, reflecting the fluctuations of the global order.

\begin{figure}[b!]
\includegraphics[width=\columnwidth]{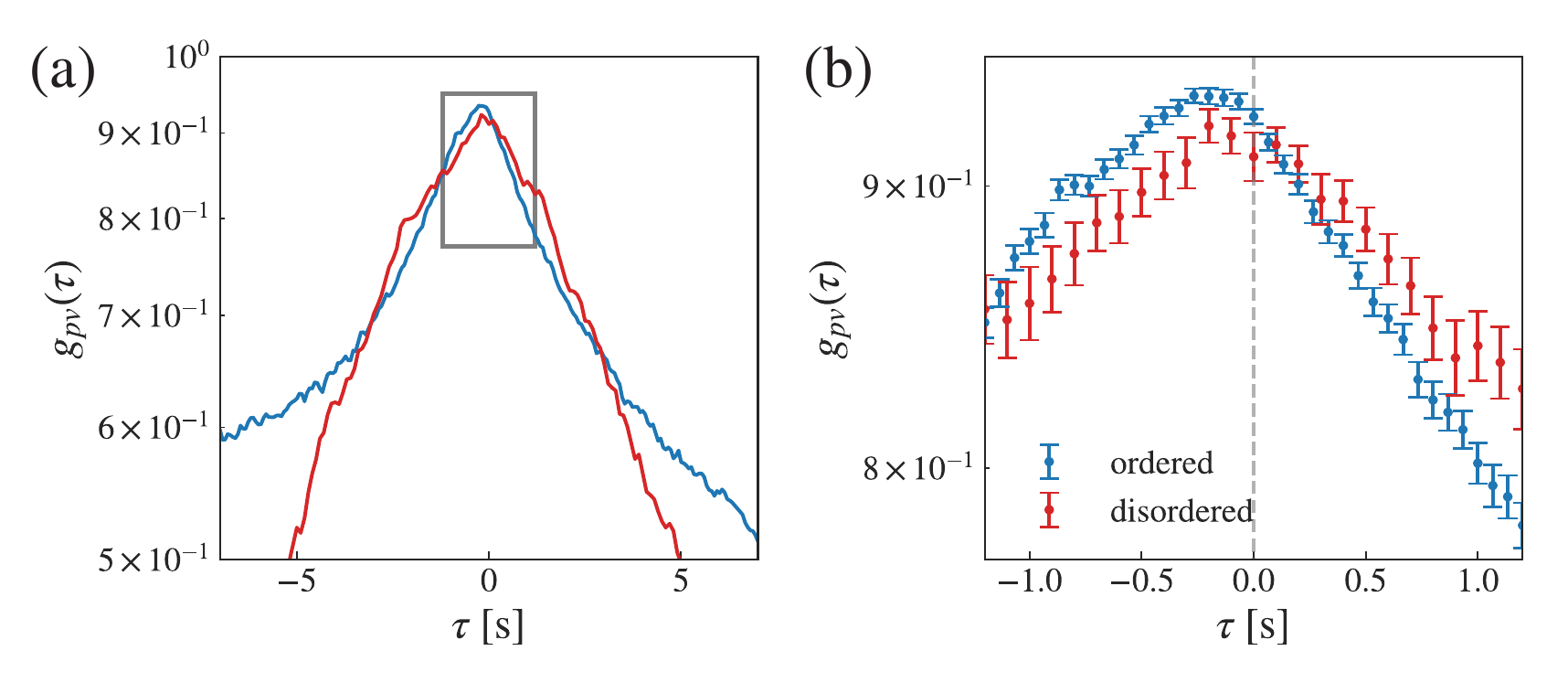}
\caption{\label{cross_corr} Polarity-velocity cross-correlation functions. 
(a) The cross-correlation function $g_{pv}(\tau)$ for Janus particles in the ordered (blue) and disordered state (red). 
(b) The same data as in (a) in a magnified view for the gray box shown in (a).
Error bars: standard error. 
The difference in error bar size is caused by the sparsity in the disordered state.}
\end{figure}

The polarity auto-correlation function for particles in the ordered state also exhibited peculiar behavior at short time scales ($\tau\sim 1$ s).
Fig.~\ref{auto_corr}(b) shows a magnified view of the behavior of $f_p(\tau)$, and it can be observed that $f_p(\tau)$ of the ordered state decays faster than that of the disordered state at short time scales.
Because this time scale is similar to the time scale of the mean free time (distance to the nearest neighbor $\sim 7~\mum$)/(average velocity $\bar{v}_o$) $\sim 0.8$ s, we speculated that this enhanced diffusion was caused by the frequent and continuous interactions among closely distanced particles in the ordered state.
To confirm this speculation, we computed the polarity-velocity cross-correlation function,
\begin{equation}
g_{pv}(\tau) = \langle\bm{n}_p(t)\cdot\bm{n}_v(t+\tau)\rangle_t,
\end{equation}
where $\bm{n}_v(t)= \bm{v}(t)/\left|\bm{v}(t)\right|$ is the unit velocity vector at time $t$ (Fig.~\ref{cross_corr}).
The assessment of $g_{pv}(\tau)$ revealed the asymmetric relation of polarity and normalized velocity in the ordered state, where polarity tended to correlate the most with the velocity of 0.2 second in the past ($\tau \sim -0.2$).
In other words, the polarity in the ordered state tended to follow the direction of the velocity of the particle with a finite time delay.
It should be noted that, although not statistically significant, $g_{pv}(\tau)$ in the disordered state showed a weak, if any, asymmetry around $\tau=0$ as well. 
These results imply that the asymmetry in $g_{pv}(\tau)$ is amplified in higher density situations.
We therefore suspect that this asymmetry is caused by electrostatic and/or hydrodynamic torques 
induced by collisions between individual particles. 
Overall, our results suggest that the particle polarities tend to follow the particle's velocity, yet local interactions between nearby particles disturb their alignment causing a time delay especially at the high density state. These effects might explain the enhanced diffusion observed in the ordered state.

\subsection{Giant number fluctuations in the ordered state}
Giant number fluctuations (GNF), where the variance of the number of particles $\Delta N^2$ grows faster than its mean $\langle N\rangle$ ($\Delta N^2\propto\langle N\rangle^{2\alpha}$ with $2\alpha>1$) in long-range ordered phases, are a hallmark of the Toner-Tu phase \cite{tonertu95, tonertu98, toner2012}. 
Since its theoretical prediction by Toner, Tu, Ramaswamy and their coworkers \cite{tonertu95, tonertu98, toner2005}, the presence of GNF has been extensively studied in numerical systems \cite{ramaswamy2003, chate2008, ginelli2010, benoit2019}. However, in addition to experimental difficulty in realizing collective motion with true long-range order, experimental detection of GNF requires delicate treatment due to many possible pitfalls such as difficulty in disentangling fluctuations arising from e.g. phase coexistence or boundary effects as critically discussed in \cite{nishiguchi2017}. Therefore, GNF as the hallmark of the Toner-Tu-Ramaswamy phase had remained undetected until recent experiments on a long-range ordered nematic phase of elongated bacteria \cite{nishiguchi2017} and then on the polar ordered phase of colloidal rollers \cite{bartolo2018}.
Here in our Janus particles system, as we have confirmed the true long-range order, it is natural to question whether the ordered state shows anomalous number fluctuations.

\begin{figure}[t!]
\includegraphics[width=\columnwidth]{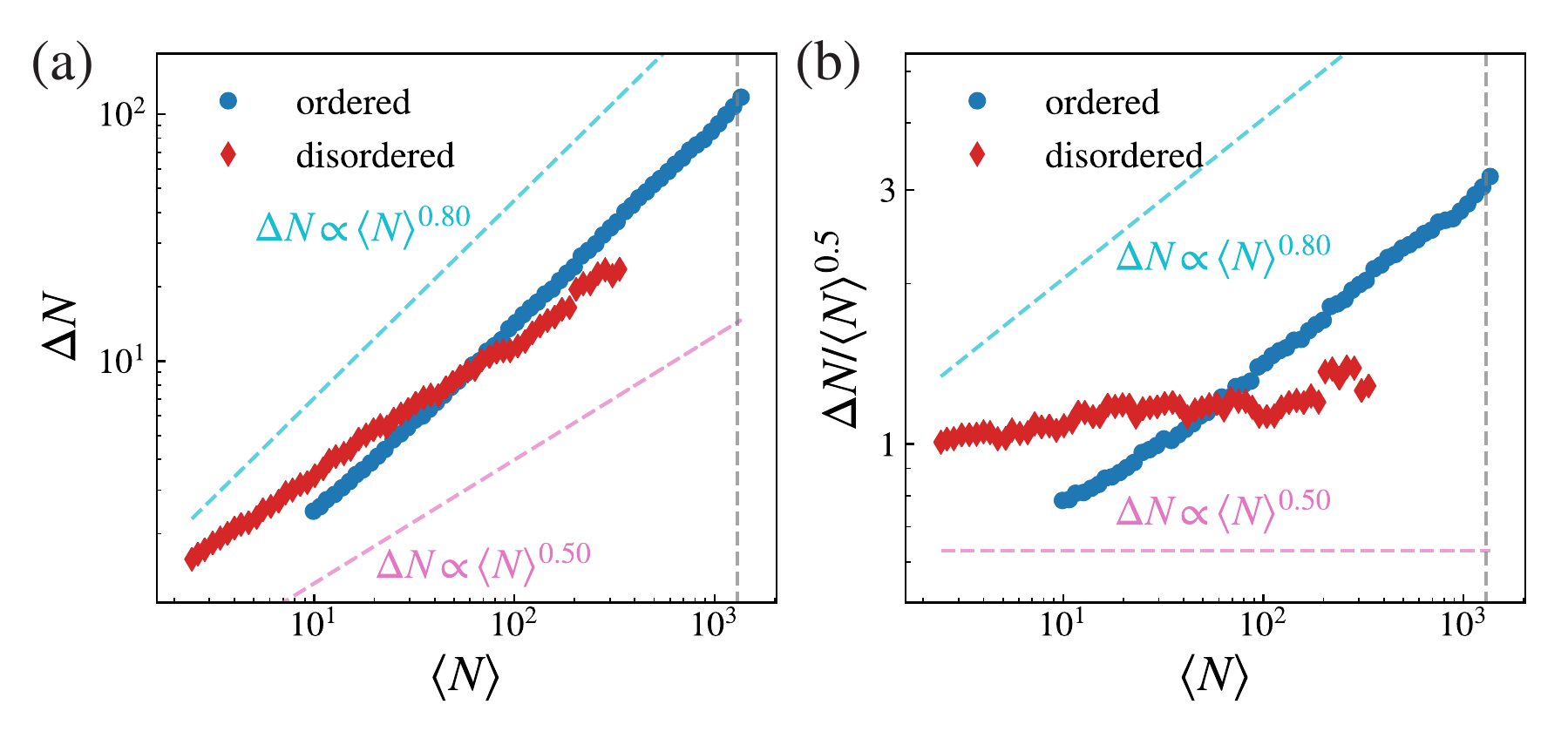}
\caption{\label{num_fluc} Scaling of number fluctuations. (a) Number fluctuations in the ordered (blue) and disordered (red) state. (b) Same data as in (a) with the number fluctuations normalized by $\langle N\rangle^{0.5}$. The cyan and magenta dashed lines show $\Delta N \propto \langle N\rangle^{0.8}$ and $\Delta N \propto \langle N\rangle^{0.5}$, respectively, as a guide to the eye. The gray dashed line indicates the scale of counterflows up to which the true long-range order persists.}
\end{figure}

To quantify number fluctuations, the mean $\langle N\rangle$ and standard deviations $\Delta N$ of the number of particles for different sizes of ROIs were calculated for statistically independent frames (Fig.~\ref{num_fluc}). 
For the disordered state, normal number fluctuations with $\Delta N \propto \langle N\rangle^{0.5}$ were observed, which is consistent with the central limit theorem.
In contrast, anomalous number fluctuations were observed in the ordered state where $\Delta N \propto \langle N\rangle^{\alpha}$ with $\alpha>0.5$. 
In fact, the estimated exponent was $\alpha=0.790(5)$ with 95\% confidence level, which is very close to the theoretical prediction $\alpha = 0.80$ \cite{tonertu95, tonertu98} and the recent numerical estimate $\alpha = 0.84$ for the Vicsek model \cite{benoit2019}. 
It is worth noting that the existence of non-trivial GNF has to be discussed under the presence of true long-range orientational order.
Indeed, there are several experimental studies that report GNF for systems that are not in the fully ordered state (for details see the supplementary material of \cite{nishiguchi2017}).
Here, we have set all ROIs to be placed at the center of the whole image so that the ROIs correspond to those that were used for the computation of the order parameter.
Accordingly, the gray dashed line in Fig.~\ref{num_fluc} indicates the length scale $S\sim 10^5\; \upmu\rm{m}^2$ at which the order parameter in the ordered phase starts to deviate from algebraic convergence to a finite value (i.e. true long-range order).
Therefore, the observation of anomalous fluctuations below this line imply that ordered Janus particles do exhibit GNF in the context of the Toner-Tu phase.

\subsection{Algebraic scars of long-range order}
When a continuous symmetry is spontaneously broken, the entire ordered phase should also be characterized by an algebraic decay of its connected correlation functions \cite{ginelli2016}. 
In the case of the Vicsek model, a power-law decay for the orientation fluctuation correlation function exemplifies the spontaneous symmetry breaking of the rotational symmetry of the system, and its exponents have been predicted by the hydrodynamic theory of Toner and Tu \cite{tonertu95, tonertu98, toner2012}.
Interestingly, a recent numerical study of the Vicsek model suggested that the ordered state could exhibit GNF exponents close to the conjectured values even though there existed a discrepancy in the exponents of the correlation functions \cite{benoit2019}.
Thus, in addition to number fluctuations, the fluctuation correlation functions are important measures to probe the statistical properties of orientational order in the system.

\begin{figure}[t!]
\includegraphics[width=\columnwidth]{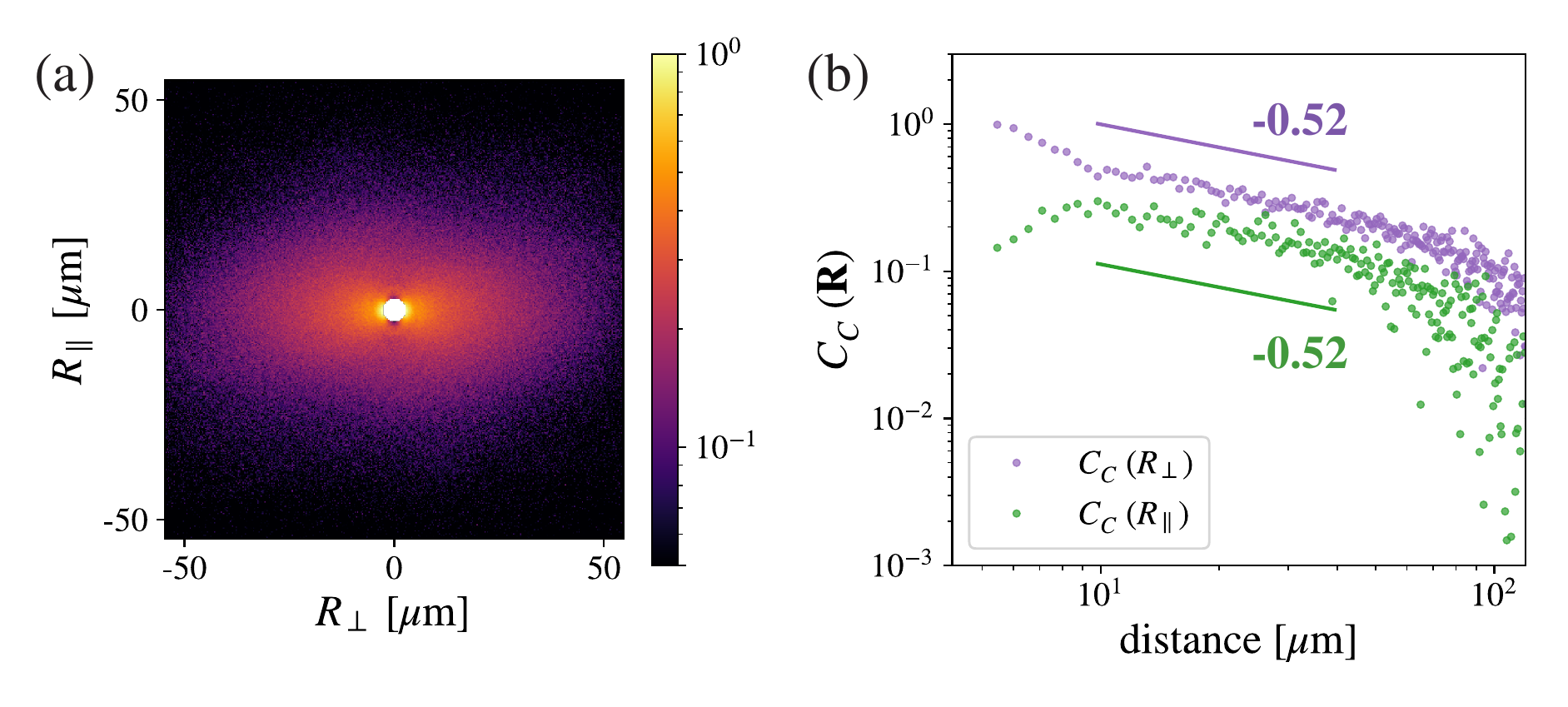}
\caption{\label{fluc_corr} Orientation fluctuation correlation function in the ordered state. 
(a) Color map of the correlation function $C_C\left(\bm{R}\right)$. 
The global mean director is set aligned with the vertical-axis ($R_\parallel$).
The region of $|\bm{R}|<6\ \mum$ were excluded from calculations due to the small statistics.
(b) Log-log plot of $C_C\left(\bm{R}\right)$ both in the transverse ($R_\perp$, purple) and longitudinal ($R_\parallel$, green) direction.
The purple and green solid lines are a guide for the eye which show a slope of -0.52.
Note, the exponents from the Toner-Tu conjecture predict an anisotropic decay of $C_C(\bm{R})\propto R_\perp^{-2/5}, R_\parallel^{-2/3}$~\cite{tonertu95, tonertu98, toner2012} and recent numerical simulations suggested $C_C(\bm{R})\propto R_\perp^{-0.62}, R_\parallel^{-0.65}$~\cite{benoit2019}.
}
\end{figure}

To address the questions of whether the orientational fluctuations of Janus particles exhibit scale-free correlations, and if so, whether they follow exponents of the Toner-Tu conjecture, we measured the orientation fluctuation correlation function,
\begin{equation}
C_C(\bm{R}) \equiv \langle \langle \delta n_{p\perp}(t,\bm{r}) \delta n_{p\perp}(t,\bm{r}+\bm{R}) \rangle_{\bm{r}} \rangle_t,
\end{equation}
where $\delta n_{p\perp}(t,\bm{r}) = \bm{n_p}(t,\bm{r}) - \langle\bm{n_{p}\rangle}$ is the deviation of a single particle polarity $\bm{n_p}$ from the global order $\langle\bm{n_{p}\rangle}$, and $\langle\cdot\rangle_{\bm{r}}$ takes an average for all $|\bm{R}|$ distanced particle pairs (see Fig.~\ref{fluc_corr}). 
Here, the global order $\langle\bm{n_{p}\rangle}$ was calculated by averaging the polarity $\bm{n_p}$ both spatially and temporally. 
Figure \ref{fluc_corr}(b) shows the decay of $C_C(\bm{R})$ in the transverse and longitudinal directions to global order.
Although there is an anisotropy in the absolute value of the correlation function, 
it decays algebraically in both directions with the same exponent close to $-0.52$.
This algebraic decay signifies the spontaneous symmetry breaking of the rotational symmetry of the system and consequent scale-free features.
However, the exponents of the decay differ from the Toner-Tu conjecture which predicts an anisotropic decay of $C_C(\bm{R})\propto R_\perp^{2\chi}, R_\parallel^{2\chi/\xi}$, where $\chi=-0.2$ and $\xi=0.6$ are the roughness and anisotropy exponent, respectively \cite{tonertu95, tonertu98, toner2012}.
This discrepancy might stem from the difference in microscopic properties including steric and hydrodynamic interactions as we will discuss in the following section.
Interestingly, recent experimental \cite{bartolo2018} and numerical \cite{benoit2019} studies for polar orientational order have both reported weak, possibly vanishing, anisotropy for the orientational fluctuation correlation function.
In addition, our analysis on the space-time correlations of density fluctuations also suggests weakly anisotropic sound modes propagating in our Janus colloidal flocks (Fig.~\ref{density_corr_prop}). 
This leads to the estimate of the dynamical exponent $z\sim1.2$ while the anisotropy exponent $\xi\sim1$, suggesting a non-diffusive but weakly anisotropic dependence that qualitatively agrees with recent numerical simulations  \cite{benoit2019}.
It should also be noted that the roughness exponent $\chi$ also takes similar but distinct exponents for theory, simulation and experiments where $\chi=-0.20$ for the Toner-Tu conjecture~\cite{tonertu95, tonertu98, toner2012}, $\chi\sim-0.31$ for numerical simulations~\cite{benoit2019}, $\chi\sim-0.38$ for active Quincke rollers~\cite{bartolo2018}, and $\chi\sim-0.26$ for our system of Janus particles.
These differences in the exponents might indicate that some of the nonlinearities that were neglected in the original calculations \cite{tonertu95,tonertu98} are relevant asymptotically, as suggested in \cite{benoit2019}.
Our measurements of $C_C(\bm{R})$ and the sound modes, along with recent numerical \cite{benoit2019} and experimental \cite{bartolo2018} studies, may pave the way for further theoretical work on collective systems.

\section{Discussion}
We studied the collective motion of active Janus particles fueled by an AC electric field. 
At high density, the Janus particles exhibited polar orientational order which allowed us to investigate and compare the statistical properties of the system with that of the Toner-Tu phase.
We have confirmed GNF within the polar orientational ordered state with an exponent similar to but different from the theoretical predictions of Toner and Tu.
In addition, an algebraic decay was observed for the orientation fluctuation correlation function which is also associated with the spontaneous symmetry breaking of the rotational symmetry of the system.
The observed exponents showed a discrepancy with the theoretical predictions and recent numerical simulations \cite{benoit2019} of the Vicsek model, suggesting a different mechanism for the emergence of long-range order.
As pointed out in \cite{nishiguchi2017} and then both experimentally and numerically verified in \cite{shi2018, tanida2020}, weak volume exclusion that allows crossing over other particles is considered a key to make collective motion of elongated objects similar to the Vicsek-style interactions and thus achieve long-range order at least for the otherwise turbulent or clustering 2D nematic systems such as bacteria and gliding microtubules \cite{nishiguchi2017,tanida2020}. In the case of polar collective motion, the weak exclusion may not be a requirement for the long-range order because the particles cannot overlap each other in our experimental setup as in granular experiments \cite{dauchot2013, ramaswamy2020}. What realizes long-range order in polar systems could be the following motion exemplified in Fig.~\ref{binary_int}, as discussed previously in e.g. mutant {\it Dictyostelium} cells \cite{shibata2019}, and/or the torque aligning the polarity to the velocity, as studied in vibrated polar disks \cite{lam2015}.

Our results might suggest that, although GNF could be observed, algebraic decay as in the Toner-Tu phases are difficult to observe in experimental systems as well as in numerical simulations.
The exponents of the algebraic decay for the orientation fluctuation correlation function differed from that observed for polar flocks of Quincke colloidal rollers \cite{bartolo2018}.
Although Janus particles and Quincke rollers are both considered as polar particles with polar interaction, the polarity acquisition mechanisms are completely different. Whereas the polarity of Janus particles is a priori defined by its metallic hemisphere, Quincke rollers acquire polarity by the spontaneous symmetry breaking of charge distribution on the particle scale.
It could be interesting to investigate whether this difference in the polarity acquisition mechanism alters the global properties of the orientationally ordered system because the latter might have additional soft modes on the particle level.

Tracking both the velocity and polarity of all particles enabled the detailed assessment of the single particle dynamics both in the disordered and ordered states.
Strikingly, the polarity auto-correlation showed a slow algebraic decay in the polar ordered state, which has never been studied even numerically. 
It could be interesting if this slow decay in polarity could be observed not only in active colloidal systems but also in collective biological matter \cite{shibata2019}. 

Our results also suggest the interplay between the polarity and velocity of the particles which might be explained by the hydrodynamic and/or the electrostatic torques between the particles.
We have shown that this interplay is amplified in the high density ordered state.
This interplay between polarity and velocity is something that cannot be observed in the polar Vicsek model where the two vectors are identical and degenerated.
A recent numerical study shows that the global properties could differ drastically by decoupling the polarity and velocity of the system \cite{benoit2018}.
Our polarity resolved realization of the collective phase of self-propelled Janus particles can contribute to bridging the understandings on the two major active matter classes: the Vicsek physics and the Active Brownian Physics.
Overall, our system of Janus particles provides a large scale statistical study for polar ordered active colloidal particles, which opens a new avenue on further theoretical and experimental works for the field of active matter.

\section{Acknowledgments}
We thank Beno\^{i}t Mahault and Hugues Chat\'{e} for insightful discussions and valuable comments on an early version of the manuscript. We also thank Tetsuya Hiraiwa, Kazumasa A. Takeuchi for fruitful discussions.
This work was supported by KAKENHI (Grant No. JP25103004). 
JI was supported by a Grant-in-Aid for the Japan Society for Promotion of Science Fellows (Grant No. JP18J21942). 
DN was supported by JSPS KAKENHI Grant Numbers JP19K23422, JP19H05800 and JP20K14426.

\bibliography{ref}
\newpage
\ 
\newpage

\newpage

\newcommand{\beginsupplement}{
        \setcounter{table}{0}
        \renewcommand{\thetable}{S\arabic{table}}
        \setcounter{figure}{0}
        \renewcommand{\thefigure}{S\arabic{figure}}
        \setcounter{equation}{0}
        \renewcommand{\theequation}{S\arabic{equation}}
        \setcounter{section}{0}
        \renewcommand{\thesection}{S\Roman{section}}
     }

\onecolumngrid
\makeatletter
\beginsupplement
\begin {center} 
	\textbf{ \large Supplemental Material for\\ ``Algebraic correlations and anomalous fluctuations in ordered flocks of\\ Janus particles fueled by an AC electric field''}\\ [.1cm]
	{Junichiro Iwasawa$^*$, Daiki Nishiguchi$^\dagger$, Masaki Sano$^\ddagger$}\\ [.1cm]
	{(Dated: \today)} \\
\end {center}

\section{Spatial Fourier transform of the density field}
\begin{figure*}[htb]
\includegraphics[width=\columnwidth]{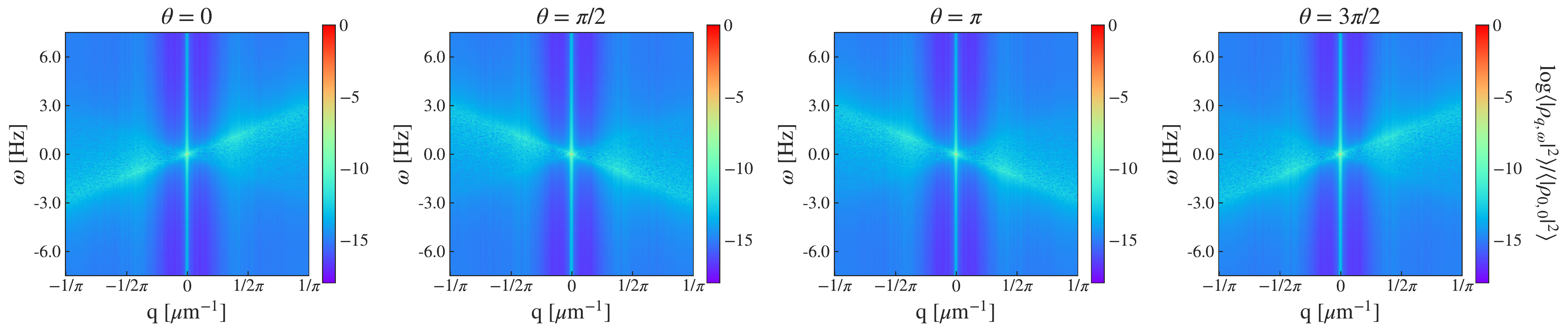}
\caption{\label{density_corr} 
Full power spectra of the density fluctuations, 
$\langle|\rho_{q,\omega}|^2\rangle / \langle|\rho_{0,0}|^2\rangle$.
The angle of the wave vector $\theta$ is defined with respect to the global order $\langle\theta\rangle$.
}
\end{figure*}

\begin{figure*}[tb]
\includegraphics[width=\columnwidth]{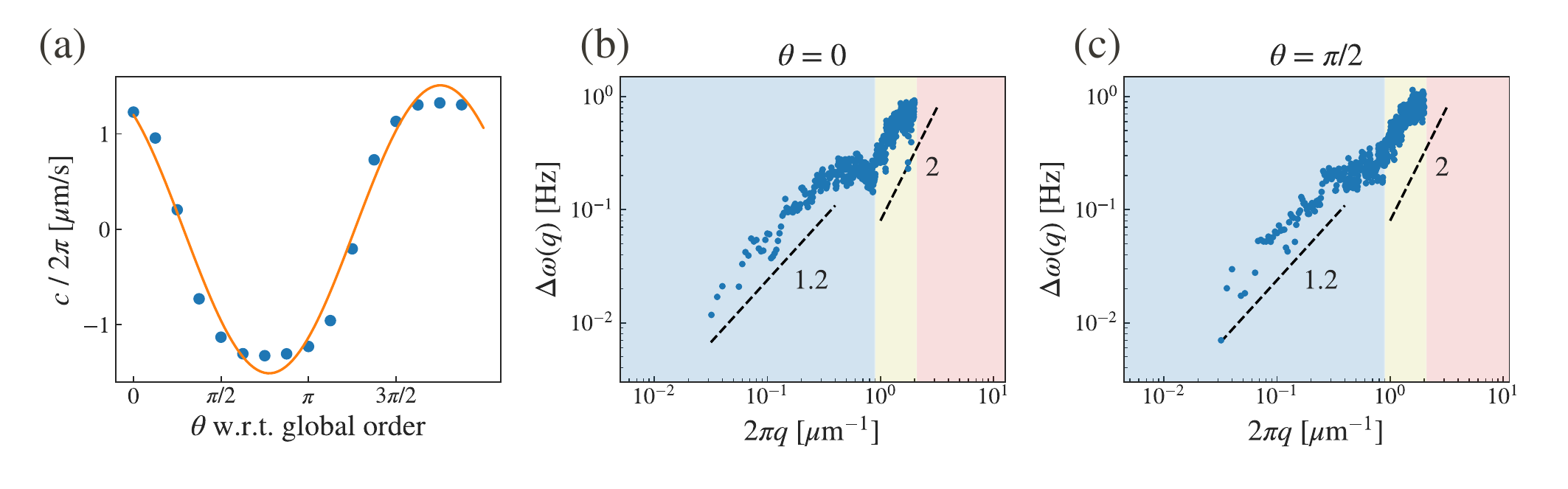}
\caption{\label{density_corr_prop} 
(a) $\theta$ dependence of the speed of sound $c(\theta)=\lim_{q\rightarrow 0}
\left[
\omega\left(\theta\right)/q
\right]$
measured from the slope at $q=0$. The blue points show $c(\theta)$ obtained from the power spectra and the orange solid line shows the fitted cosine curve: $a\cos{(b\theta+\delta)}$ with $a=9.5$, $b=1.0$, and $\delta=0.65$.
(b), (c) $\Delta\omega(\bm{q})$ characterized as the full width at half maximum of the fitted Lorentzian for the Fourier transformed density correlation function as functions of $\bm{q}$ in the longitudinal (b) and transverse (c) direction with respect to the global order. The black dashed line with a slope: 1.2, 2 is a guide for the eye. The red, blue and yellow areas correspond to length scales smaller than the particle's diameter  ($<3~\upmu$m), length scales larger than the typical interparticle length in the ordered state ($>7~\upmu$m), and length scales in between, respectively.
}
\end{figure*}

To investigate the sound modes in the ordered state of Janus particles, we computed the power spectra using the particle coordinates \cite{bartolo2018}.
The spatial Fourier transform of the density field is defined as,
\begin{equation}
    \rho_q(t) = \sum_j e^{iq\left[x_j(t)\cos\theta + y_j(t)\sin\theta\right]},
\end{equation}
where $x_j(t)$, $y_j(t)$ are the instantaneous particle coordinates, $\sum_j$ takes an average of all particles and $q(\cos\theta, \sin\theta)$ is the wave vector making an angle $\theta$ with the global order $\langle\theta\rangle$. 
Here, the global order is defined as the mean polarity of all particles averaged through all frames. 
The power spectra $\left|\rho_{q,\omega}\right|^2$ (Fig.~\ref{density_corr}) were obtained by performing time Fourier transformations to the two-time autocorrelations of the density fields $\langle\rho_q(t)\cdot\rho_q^{\dagger}(t+\tau)\rangle_t$. 
The speed of sound modes
$c(\theta)=\lim_{q\rightarrow 0}
\left[
\omega\left(\theta\right)/q
\right]$
were obtained from the positions of the peaks in the dispersion relations given by the power spectra (Fig.~\ref{density_corr_prop}).
Here, the positions and widths $\Delta\omega$ of the peaks as a function of $q$ in the power spectra were determined by fitting the $\left|\rho_q(\omega)\right|^2$ curve using a Lorenztian function, $f(x)=\frac{a}{\pi\gamma\left[1+\left(\frac{x-x_0}{\gamma}\right)^2\right]}+b$, where $a,\gamma,x_0,b$ are fitting parameters.

The $\theta$ dependency of $c(\theta)$ could be fitted by a cosine curve: $a\cos{(b\theta+\delta)}$ with $a=9.5$, $b=1.0$, and $\delta=0.65$, respectively (Fig.~\ref{density_corr_prop}(a)). 
The speed of sound $c(\theta)$ is comparable to the mean velocity of the particles in the ordered state ($v_o\sim 8.5\upmu$m/s) which indicates that the observed density fluctuations are caused by the advection of the particles. In addition, sound modes of particles that move in the same direction with a uniform velocity exhibit a sinusoidal dependency against $\theta$, which again suggests that the sound modes observed in our system originate from the advection in the system.

We have also plotted the peak widths $\Delta\omega$, characterized as the full width at half maximum (FWHM) of the fitted Lorenztian, as a function of $\bm{q}$ (Fig.~\ref{density_corr_prop}(b), (c)).
Although the statistical noise in our system kept us from observing the two mixed sound modes as predicted by Toner and Tu, we were still able to observe the algebraic behaviors underlying the sound modes.
For both longitudinal and transverse directions, $\Delta\omega(\bm{q})$ showed a slope close to 2 where $3~\upmu\textrm{m}< 1/|\bm{q}| <7~\upmu\textrm{m}$, indicating that the sound modes are subject to diffusive damping.
On the other hand, $\Delta\omega(\bm{q})$ showed a slope smaller than 2 at length scales larger than typical interparticle lengths $7~\upmu\textrm{m}< 1/|\bm{q}|$ which suggests the existence of long-range correlations that suppress diffusive damping.
Interestingly, $\Delta\omega(\bm{q})$ showed similar exponents for both longitudinal and transverse directions which is consistent with the results of $C_C\left(\bm{R}\right)$ (Fig.~\ref{fluc_corr}), indicating weak anisotropic correlations with respect to global order.
Note, the theoretical predictions given by Toner and Tu are $\Delta\omega\sim q_{\parallel}^{z/\xi}, q_{\perp}^{z} (q_{\parallel,\perp}\rightarrow0)$, where $z=1.2$, $\xi=0.6$ are the dynamical and anisotropy exponent, respectively.

\section{Calculation of the polar order parameter}
\label{sup_calculation_order_parameter}
\begin{figure*}[thb]
\includegraphics[width=0.8\columnwidth]{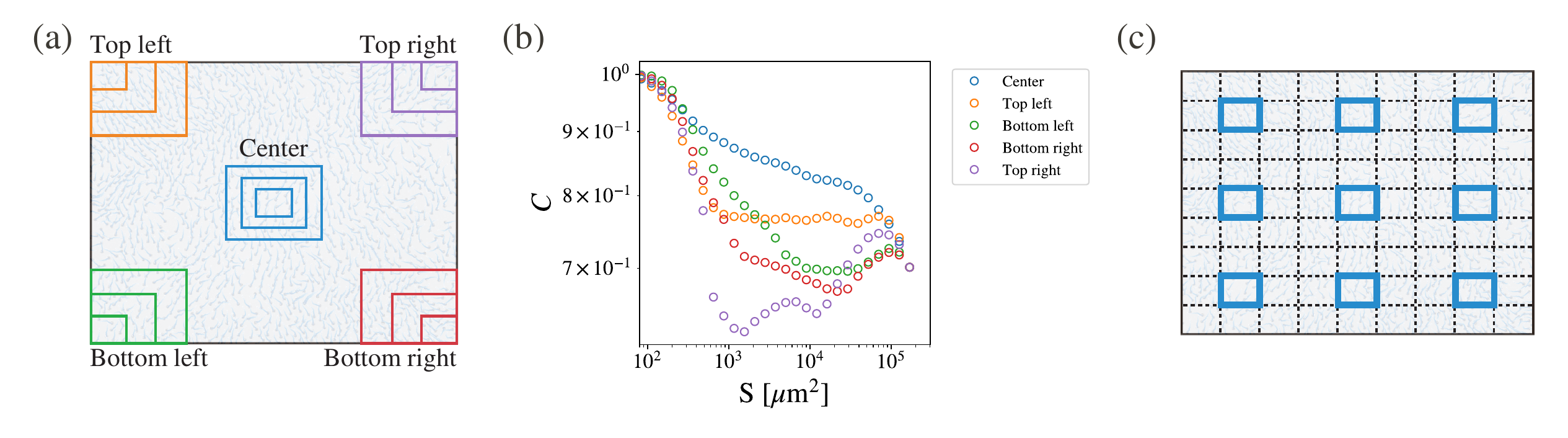}
\caption{\label{sup_sampling_bias} 
(a) A schematic picture showing how we have sampled ROIs for calculating the relation of $C$ and $S$ in (b). 
The ROIs were sampled from the center (blue), top left (orange), bottom left (green), bottom right (red) and the top right (purple) of the whole field.
(b) The plot for the polar order parameter $C$ in the ordered state vs $S$ the area of the ROI for ROIs sampled from different locations of the whole view.
(c) A schematic picture showing how we have sampled ROIs for calculating the polar order parameter $C$ in Fig.~\ref{order_param}.
We have divided the whole view to grids and sampled a ROI in every three grids (both in the horizontal and longitudinal direction) to prevent spatial correlations within the sampled ROIs.
}
\end{figure*}

To assess the relation between $C$ and $S$ when the center of ROI is taken in different places compared to that when the ROI is fixed to the center of the whole field, we have calculated $C$ with ROIs fixed to the center, top left, bottom left, bottom right, and the top right of the whole field, respectively (Fig.~\ref{sup_sampling_bias}(a)(b)).
This analysis indicated that using ROIs only from a certain location gives systematically erroneous results for $C$.
Therefore for every ROI size, we have divided the whole view into grids and sampled ROIs from various locations to reduce systematical errors for $C$ (Fig.~\ref{sup_sampling_bias}(c)). 
Here, we have sampled a ROI in every three grids (both in the horizontal and longitudinal direction) to prevent spatial correlations among the sampled ROIs.
This results in a total number of 36 ROIs to average, for example, when $S\sim500~\upmu\textrm{m}^2$.
The frames were also sampled every 100 frames ($\sim$ every six seconds in the ordered state) to prevent temporal correlations.
The results for $C$ are shown in Fig.~\ref{order_param}.
$C$ based on ROIs sampled from the whole view shows an algebraic decay to an asymptotic value $C_\infty=0.74$ with a slope of $S^{-0.53}$ (Fig.~\ref{order_param}(b)). 
The result where $C$ based on ROIs sampled from the whole view exhibits a slower decay than a power law, indicates that the asymptotic behavior of $C$ is robust.

\section{Auto-correlation functions}
\begin{figure*}[hbt]
\includegraphics[width=0.3\columnwidth]{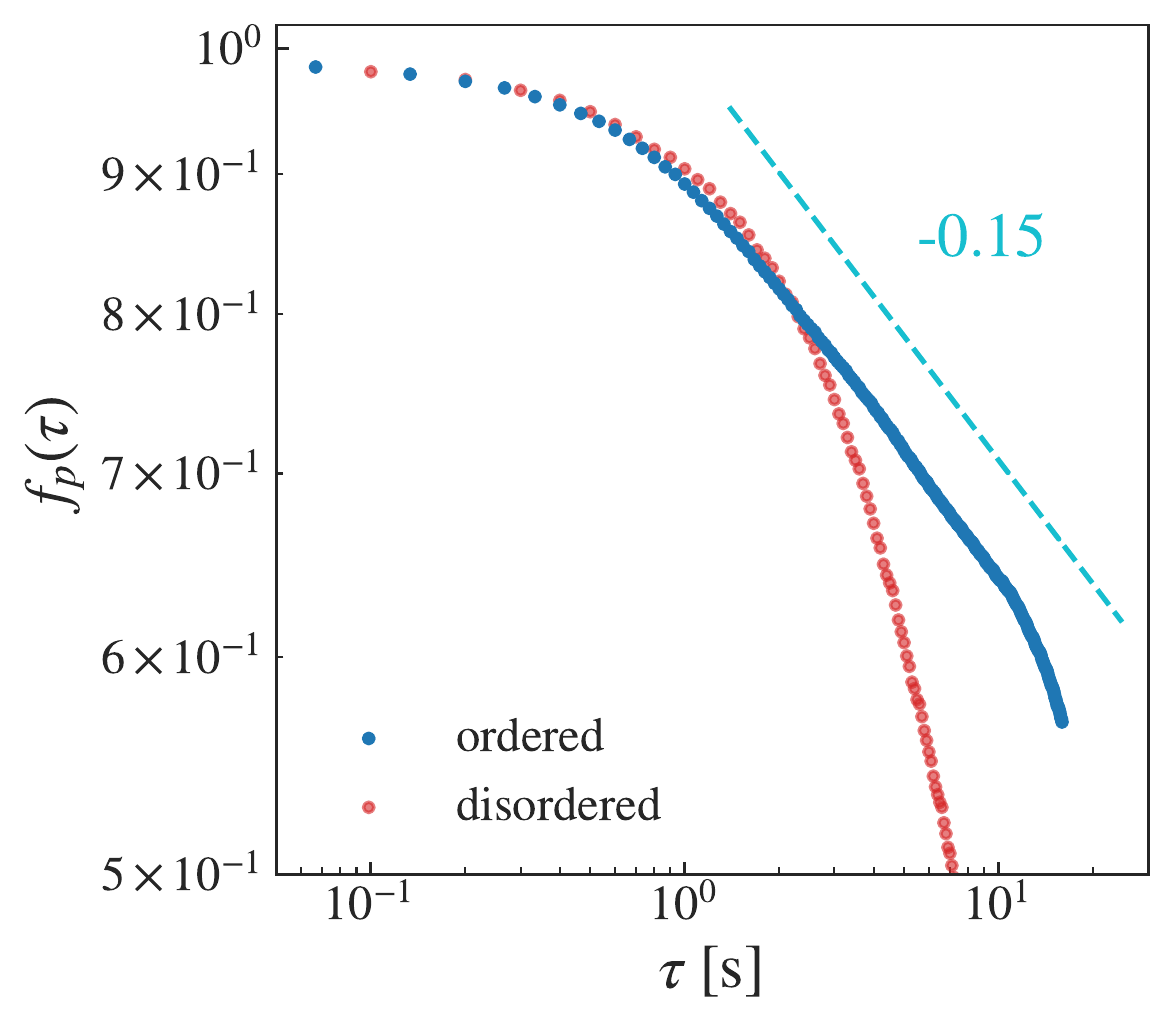}
\caption{\label{sup_autocorr} Polarity auto-correlation functions $f_p(\tau))$ in the ordered and disordered state (log-log). The same data as in Fig.~\ref{auto_corr}(a) in a log-log plot is shown.}
\end{figure*}

\begin{figure*}[thb]
\includegraphics[width=0.3\columnwidth]{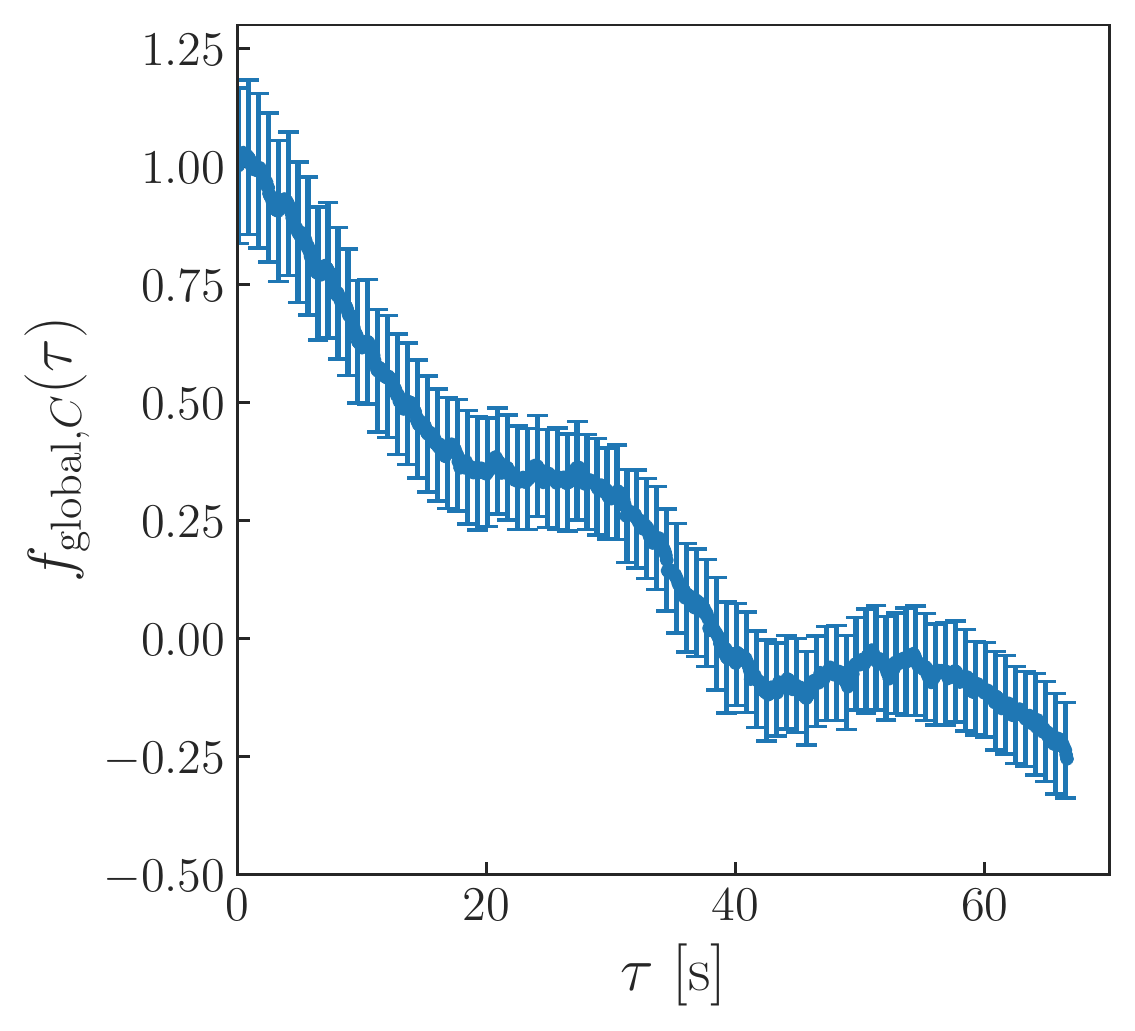}
\caption{\label{sup_global_order} Auto-correlation of the polarity fluctuations from global order 
$f_{\rm global,C}(\tau)=\langle\Delta\tilde{\bm{n}}_p(t)\cdot\Delta\tilde{\bm{n}}_p(t+\tau)\rangle_t$. 
Error bars show the standard error.}
\end{figure*}

The polarity auto-correlation function $f_p(\tau)$ in a log-log plot is shown in Fig.~\ref{sup_autocorr}. 
A slow power-law-like decay for $\tau > 1$~s with a slope close to $-0.15$ can be observed for $f_p(\tau)$ in the ordered state.
The auto-correlation function of the polarity fluctuations from global order 
$f_{\rm global,C}(\tau)=\langle\Delta\tilde{\bm{n}}_p(t)\cdot\Delta\tilde{\bm{n}}_p(t+\tau)\rangle_t$,
where
$\Delta \tilde{\bm{n}} _{p}(t) = \langle\bm{n}_p - \langle\bm{n}_p\rangle_{t,\textrm{ordered}}\rangle_{\textrm{ordered}}$,
is shown in Fig.~\ref{sup_global_order}.
$f_{\rm global,C}(\tau)$ shows an oscillation with a period of $\sim 20$ s while it decays. 
This is consistent with the time scale of the time needed for the particles to propel through the ordered area $\bar{v}_o \times 20$ s $\sim 170\ \upmu$m which is close to the length scale where the order parameter deviates from true long-range order.

\section{Polarity detection via ImageJ}
\begin{figure*}[!bt]
\includegraphics[width=\columnwidth]{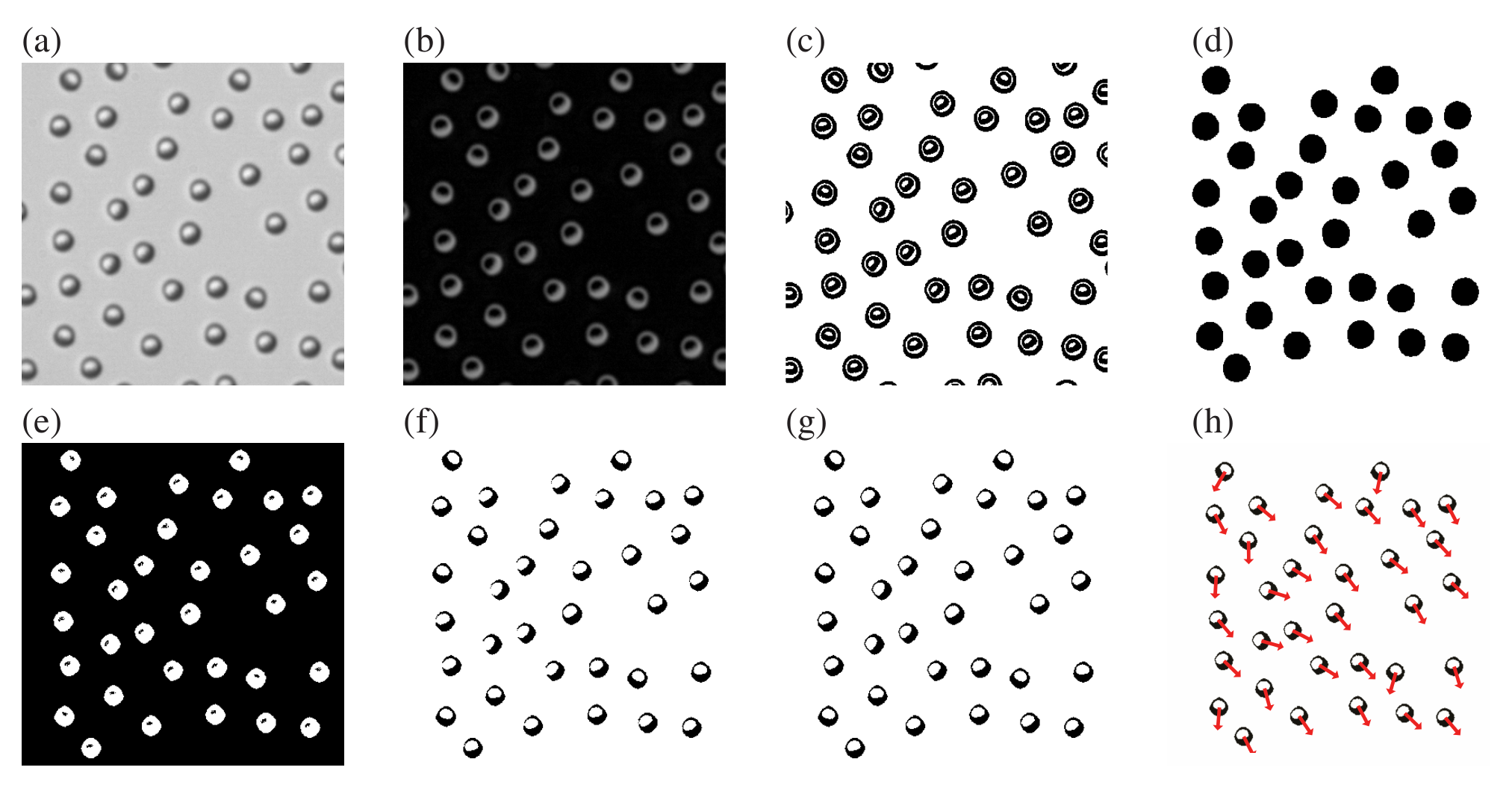}
\caption{\label{imagej} An example of the image analysis procedure using ImageJ. 
(a) The raw image of Janus particles ($397\times397$ pixels, $57\mum\times57\mum$). 
(b) Image of Janus particles after inverting and subtracting the background using the rolling ball method from the original image. 
(c) The edges detected from the image in (b). 
(d) Binary masks of the whole particle. 
(e) Multiplied image of the inverted image of particles (b) and the masks (d). 
(f) The thresholded image of (e). 
(g) The OR operated image of the outlines of (d) and (f). 
(h) Red arrows indicate the orientation of each particle from the difference in the coordinates of the centroid and the center of mass of luminance.
}
\end{figure*}

In order to obtain the polarity of each particle, we have applied image analysis methods
using ImageJ \cite{imagej}. 
Here we give a description of the whole image analysis procedure in order to obtain the orientation which is defined by the metal hemisphere. 
The main idea is to binarize the images of Janus particles and obtain the orientation of each particle by taking the difference of the center of mass (intensity) and the centroid. In Fig.~\ref{imagej}, we show how the processed image will look like when we perform the following procedure on the raw experimental image (Fig.~\ref{imagej}(a)).
\begin{enumerate}
    \item We first inverted the whole image (Edit$>$Invert) and subtracted the background using the rolling ball method (Edit$>$Invert, Process$>$Subtract Background).
    Since the radius of the particle was $\sim 11$ pixels, we set the rolling ball
    radius to 20.0 pixels (Fig.~\ref{imagej}(b)).
    
    \item We next detected the edges (Process$>$Find Edges) and binarized the image (Image$>$Adjust$>$Threshold). The thresholding was done by auto
    thresholding (Fig.~\ref{imagej}(c)).
    
    \item In order to make the particle detection easier and more precise, we dilated
    the edges (Process$>$Binary$>$Dilate) and filled in the holes of the particle (Process$>$Binary$>$Fill Holes).
    
    \item We detected the particles using the ``Analyze Particles'' program in ImageJ (Analyze$>$Analyze Particles).
    In order to neglect the noise in the image (e.g. small fragments of metal), we only
    detected particles which were larger than 400 pixels. By using ``Analyze Particles''
    we obtained the masks of each particle (Fig.~\ref{imagej}(d)).
    
    \item Since the generated masks were slightly larger than the real particles, the images were
    eroded for four times (Process$>$Binary$>$Erode).
    
    \item Using the image calculator, we multiplied the inverted image of the particles and the eroded mask (Process$>$Image Calculator$>$
    Multiply, Fig.~\ref{imagej}(e)). The obtained image was thresholded thereafter (Image$>$Adjust$>$Threshold, (f)).
    
    \item We obtained the outlines of the particles (Process$>$Binary$>$Outline) and added the outlines with the
    thresholded image (Process$>$Image Calculator$>$OR, Fig.~\ref{imagej}(g)). The center of mass of image intensity was calculated from this image.
    
    \item By taking the difference of the position vectors of the centroid and the center
    of mass of image intensity, we could obtain the orientation of each particle. The coordinates of the centroid and the center of mass were calculated by ImageJ (Fig.~\ref{imagej}(h)).
\end{enumerate}

\section{Details of experimental procedures}

We used Trackpy \cite{trackpy} for linking trajectories.
The instantaneous velocities $\bm{v}$ of the particles were computed by taking the difference of the centroid:
$\bm{v}(t) = \bm{x}_{c_i}(t+1) - \bm{x}_{c_i}(t-1)$,
where $t$ denotes the frame number and $\bm{x}_{c_i}(t)$ is the coordinate for the centroid of particle $i$ at frame $t$. We used the instantaneous velocities for all particles for every 100 frames for computing the velocity distribution in Fig.~\ref{ordered_janus}(c)(d).
For the computation of the auto-correlation function $f_v(\tau)$, the velocity was exceptionally defined as,
$\bm{v}(t) = \bm{x}_{c_i}(t+3) - \bm{x}_{c_i}(t)$,
in order to reduce noise in the measurements by using a larger window.
It should be noted that, we used the symmetric definition $\bm{v}(t) = \bm{x}_{c_i}(t+1) - \bm{x}_{c_i}(t-1)$ for the computation of the cross-correlation function $g_{p,v}(\tau)$ to avoid confusion in the interpretation.

\begin{figure*}[hbt]
\includegraphics[width=0.88\columnwidth]{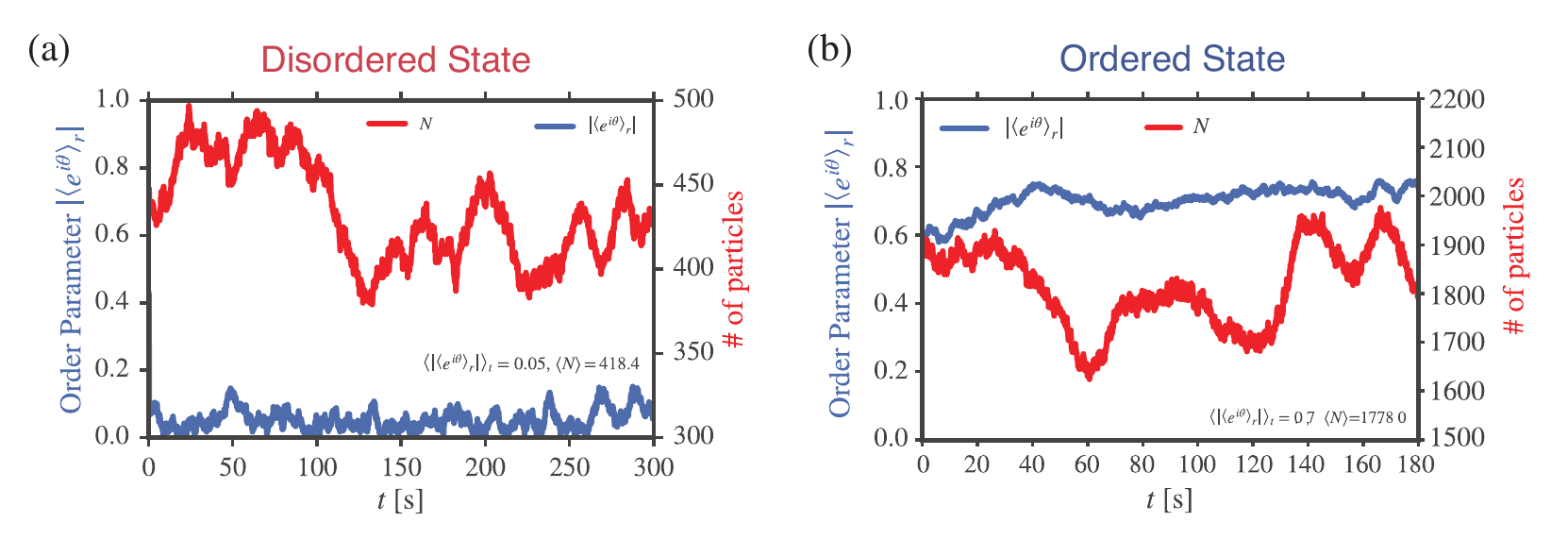}
\caption{\label{timeseries} Time series for the total number of particles $N$ and the polar  order parameter $\left|\langle e^{i\theta}\rangle_r\right|$. (a) Time series for the disordered state. (b) Time series for the ordered state.}
\end{figure*}

The order parameter in Fig.~\ref{order_param} was computed using the particle polarities detected by ImageJ. The time series of the total number of particles in the whole field of view and the polar order parameter for the whole view are given in Fig.~\ref{timeseries}.

For inferring binary interactions in Fig.~\ref{binary_int}, the locations of particle pairs within 6 $\upmu$m and 30 $\upmu$m distance were extracted for statistically independent frames which were 25 frames ($=2.5$ s) apart.

\section{Movie description}
\begin{description}
\item[Supplemental Movie 1] (1\_Disordered.mp4)\\
Janus particles in the orientationally disordered state. The particles were exposed to a 1000 kHz, $16~{{\rm V}_{{\rm pp}}}$ AC electric field. The movie shows the first 1800 frames out of the 3600 frames used for binary collision and other statistical analysis. The resolution is scaled down to $640\times512$ from the original $3000\times2400$ to reduce the file size. The movie is played at three times the real speed.

\item[Supplemental Movie 2] (2\_Disordered\_crop.mp4)\\
The same experiment as Supplemental Movie 1 with a cropped view of $562\times382$ pixel view for the 15th -- 55th frames. The original resolution is kept to ensure the visibility of each particle's polarity. The movie is played at the real speed.

\item[Supplemental Movie 3] (3\_Ordered.mp4)\\
Janus particles in the orientationally ordered state. The particles were exposed to a 1000 kHz, $16~{{\rm V}_{{\rm pp}}}$ AC electric field. The movie shows the first 1800 frames out of the 3180 frames used for statistical analysis. The resolution is scaled down to $640\times512$ from the original $3000\times2400$ to reduce the file size. The movie is played at two times the real speed.

\item[Supplemental Movie 4] (4\_Ordered\_colored.mp4)\\
Janus particles in the orientationally ordered state with colored by their deviation from global order. The movie shows the 1500th -- 3000th frames out of the 3180 frames used for statistical analysis. The colors show the polarity deviation for each particle from the global order in radian. The movie is played at four times the real speed.
\end{description}

\end{document}